\documentclass[journal, onecolumn]{IEEEtran}

\IEEEoverridecommandlockouts
\usepackage{cite}
\usepackage{theorem}  
\usepackage{amsmath,amssymb,amsfonts}
\usepackage{algorithmic}
\usepackage{graphicx}
\usepackage{textcomp}
\usepackage{xcolor}
\usepackage[ruled,vlined,linesnumbered]{algorithm2e}
\usepackage{float}
\def\BibTeX{{\rm B\kern-.05em{\sc i\kern-.025em b}\kern-.08em
    T\kern-.1667em\lower.7ex\hbox{E}\kern-.125emX}}

\usepackage{pgfplots}
\usepgfplotslibrary{dateplot}
\pgfplotsset{compat=1.8}

\allowdisplaybreaks

\newcommand{\be}[1]{\begin{equation}\label{#1}}
\newcommand{\ee}{\end{equation}}

\newcommand{\bc}{\begin{center}}
	\newcommand{\ec}{\end{center}}

\newcommand{\floor}[1]{\lfloor{#1}\rfloor}

\newcommand{\ceil}[1]{\lceil{#1}\rceil}

\renewcommand{\leq}{\leqslant}

\renewcommand{\geq}{\geqslant}

\newcommand{\R}{\mathbb{R}}

\newcommand{\N}{\mathbb{N}}

\DeclareMathOperator*{\argmin}{arg\,min}

\newcommand{\Cref}[1]{Co\-rol\-la\-ry\,\ref{#1}}

\theoremstyle{plain} \theorembodyfont{\normalfont\slshape}

\newtheorem{thm}{Theorem$\!$}
\newenvironment{theorem}{\begin{thm}\hspace*{-1ex}{\bf.}}{\end{thm}}

\newtheorem{prop}[thm]{Proposition$\!$}

\newtheorem{lem}[thm]{Lemma$\!$}
\newenvironment{lemma}{\begin{lem}\hspace*{-1ex}{\bf.}}{\end{lem}}

\newtheorem{cor}[thm]{Corollary$\!$}

\newtheorem{defi}[thm]{Definition$\!$}

\newtheorem{cl}[thm]{Claim$\!$}

\theorembodyfont{\normalfont}

\newtheorem{exam}{Example$\!$}

\newtheorem{remrk}{Remark$\!$}

\newtheorem{const}{Construction$\!$}

\newcommand{\al}[1]{\begin{align}#1\end{align}}

\newcounter{numcount}
\newcommand{\eqnum}{\stackrel{(\roman{numcount})}{=}\stepcounter{numcount}}

\newcommand{\Ber}{{\rm Ber}}
\newcommand{\Geom}{{\rm Geom}}

\newif\iflong
\longfalse

\newif\ifdraft
\drafttrue

\newcommand{\longversion}[1]{
\iflong
#1
\else
\fi
}

\newcommand{\iscomment}[1]{
\ifdraft
{\color{blue} \bf{{{{IS --- #1}}}}}
\else
\fi
}

\newcommand{\rdcomment}[1]{
\ifdraft
{\color{blue} \bf{{{{RD --- #1}}}}}
\else
\fi
}

\newcommand{\kmcomment}[1]{
\ifdraft
{\color{orange} \bf{{{{KM --- #1}}}}}
\else
\fi
}

\usepackage[colorinlistoftodos]{todonotes}

\usepackage[numbers,sort&compress]{natbib}

\newcommand{\bluechange}[1]{{\leavevmode\color{black}{#1}}}

\begin{document}
\pagestyle{plain}

\title{
Private DNA Sequencing: \\
Hiding Information in Discrete Noise \\
\thanks{
The work of Kayvon Mazooji and Ilan Shomorony was supported in part by the National Science Foundation (NSF) under grants \mbox{CCF-2007597} and \mbox{CCF-2046991}.

Kayvon Mazooji, Roy Dong, and Ilan Shomorony are with the Department of Electrical and Computer
Engineering, University of Illinois at Urbana–Champaign (UIUC), Urbana,
IL 61801 USA 
(e-mail: \mbox{mazooji2@illinois.edu}, \mbox{roydong@illinois.edu}, \mbox{ilans@illinois.edu}).

An earlier version of this article was presented in
part at the 2020 Information Theory Workshop (ITW) \cite{Mazooji}.}
}

\author{
\IEEEauthorblockN{Kayvon Mazooji, Roy Dong, Ilan Shomorony}
}
\maketitle

\begin{abstract}
When an individual's DNA is sequenced,
sensitive medical information becomes available to the sequencing laboratory.
A recently proposed way to hide an individual's genetic information is to mix in DNA samples of other individuals. We assume that the genetic content of these samples is known to the individual but unknown to the sequencing laboratory. Thus, these DNA samples act as ``noise'' to the sequencing laboratory, but still allow the individual to recover their own DNA samples afterward.
Motivated by this idea, 
we study
the problem of hiding a binary random variable $X$ (a genetic marker) with the additive noise provided by mixing DNA samples, using mutual information as a privacy metric.
This is equivalent to the problem of finding a worst-case noise distribution for recovering $X$ from the noisy observation
among a set of feasible discrete distributions.
We characterize upper and lower bounds to the solution of this problem, which are empirically shown to be very close.
The lower bound is obtained through
a convex relaxation of the original discrete optimization problem, and yields a closed-form expression.
The upper bound
is computed via a greedy
algorithm for selecting the mixing proportions.
\end{abstract}



\section{Introduction}
\label{sec:intro}

Advances in DNA sequencing technologies have led to the generation of human genetic data at an unprecedented rate~\cite{astronomical}. 
This offers exciting prospects for biomedical research, and recent studies have leveraged the genetic data of hundreds of thousands of individuals to identify genetic markers associated with many traits and diseases \cite{cad,diabetes,alzheimer,cirulli_rare}.


Genetic testing for disease predisposition \cite{diseaserisk} and popular direct-to-consumer genomics services \cite{directtoconsumer,directtoconsumer2} can provide us with  important and actionable information about our health. 
However, these services require the submission of a blood or saliva sample, making an individual's entire DNA available to the testing center.
This raises significant privacy concerns regarding genetic data \cite{erlich}, particularly with respect to the potential use of this information by insurance companies \cite{23andmepharma}.

Given the potential privacy risks of DNA sequencing, an important question is whether it is possible to alter a physical DNA sample prior to submitting it to a laboratory, in order to ``hide'' some of its genetic information.
One possible way to alter a sample could be to 
\emph{mix} it with the DNA of other individuals.
Upon sequencing, the lab would then observe 
a mixture of 
the data from the different samples, 
which would hinder its ability to retrieve individual genetic variants.

\begin{figure}[b!]
	\centering
 	\includegraphics[width=\linewidth]{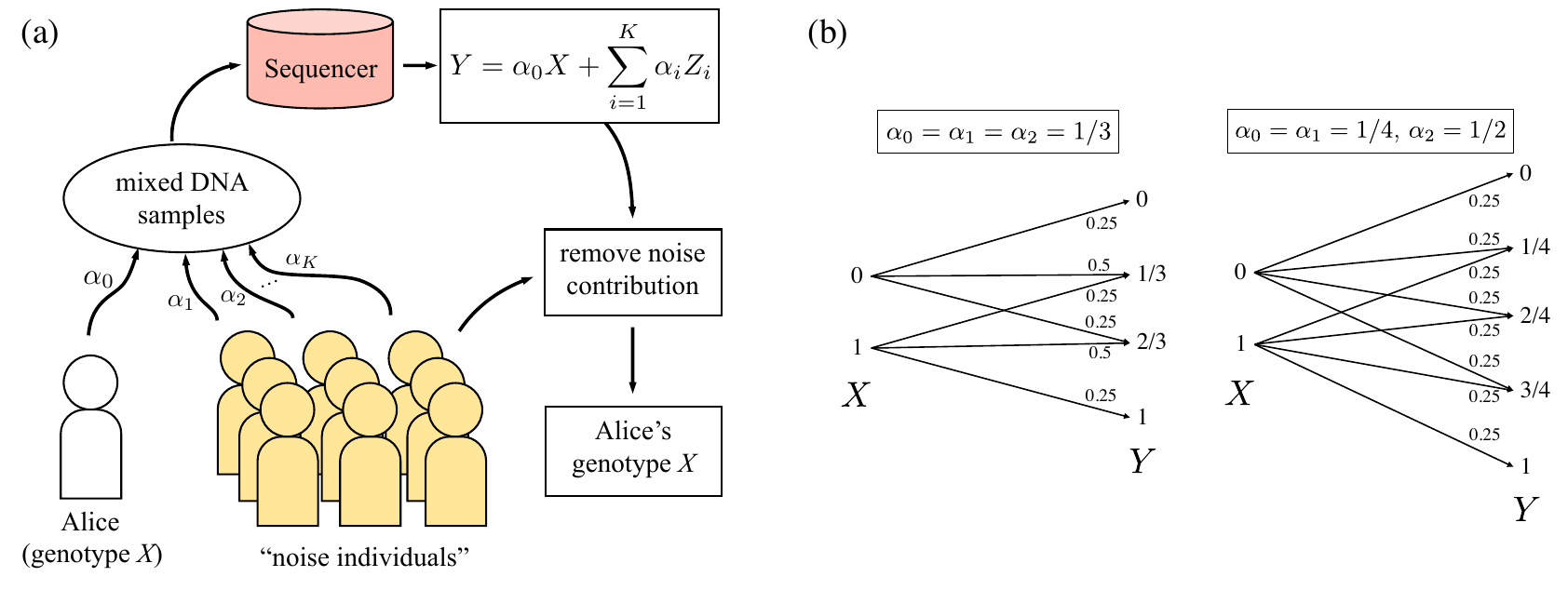}
	\caption{
	(a) In order to hide her genotype $X$ at a given locus $s$, Alice mixes her DNA sample with that of $K$ individuals in amounts $\alpha_1,...,\alpha_K$.
	Upon receiving the sequencing data from the lab, Alice can remove the contribution from the ``noise individuals'' (whose genotype at $s$ is known) to recover $X$.
    (b)~Depending on the choice of the mixing coefficients $\alpha_0,\dots,\alpha_K$, very different channels are created between $X$ and the lab's observation $Y$.
  For $K=2$, the scheme $\alpha_0 = \alpha_1 = \alpha_2 = 1/3$ leads to the channel on the left, while the scheme $\alpha_0 = \alpha_1 = 1/4$, $\alpha_2 = 1/2$ leads to the channel on the right. 
  Notice that even the output alphabet changes as a function of the mixing coefficients.
	}
	\label{fig:Overview}
\end{figure}

The general idea of mixing samples to attain genetic privacy
was proposed in \cite{Maddah-Ali} (and later extended in \cite{Maddah-Ali2}).
Suppose Alice wants to have her DNA
sequenced and has at her disposal the
DNA samples of $K$ other people
who \emph{already know} their DNA sequence \bluechange{and are willing to share this information with her}.
Alice can then mix all $K+1$ DNA samples and send them to the sequencing lab.
From the lab's perspective, the DNA of the $K$ additional individuals plays the role of noise, impairing the lab's ability to recover Alice's DNA.
However, upon receiving the sequencing data back, 
the contribution of the ``noise individuals'' can be removed, and Alice can recover her DNA sequence information.
This approach is illustrated in Figure~\ref{fig:Overview}(a). 
\bluechange{We point out that the mixing of individuals' biological samples is also present in the literature on group testing, where the goal is to efficiently identify infected individuals \cite{Aldridge2019}. 
Hence, this framework can be seen as utilizing a form of quantitative group testing~\cite{wang2017optimal} as a mechanism for privacy.}





Motivated by this setting, 
we study how to optimally choose the proportions of each individual's DNA sample in the mixture in order to maximize the privacy achieved.
\bluechange{We observe that varying the proportions with which the individuals are mixed in can drastically affect the level of privacy from the sequencing laboratory, and previous works \cite{Maddah-Ali, Maddah-Ali2} do not study this phenomenon.}
We focus on a single \emph{biallelic site} $s$ on the genome; i.e., a location on the human genome that admits two possible alleles, and can thus be modeled as a single variable $X \in \{0,1\}$.  It should be noted that the vast majority of genome locations where a variant has been observed are biallelic \cite{Casci}.
A biallelic site could model, for example, the presence of the mutation on the BRCA2 gene that increases the likelihood of breast cancer \cite{brca2} and many other disease genetic markers.
\bluechange{Previous works focus on understanding the privacy of a fixed scheme that sequences multiple unknown genomes at once for multiple biallelic sites.  In contrast, our work seeks to understand the \emph{optimal} privacy achievable at a single biallelic site for a single unknown genome as an initial step.  
This yields a simply stated, yet rich information-theoretic problem which we attempt to better understand.}


\longversion{
\iscomment{mention diploid in long}
}


In order to hide her genotype $X$ from the sequencing lab, Alice mixes into her sample the samples of $K$ individuals using proportions $\alpha_0,\alpha_1,...,\alpha_K$, where $\sum_{i=0}^K \alpha_i = 1$.
We model the lab's observation of site $s$ as
$Y = \alpha_0 X + \sum_{i=1}^K \alpha_i Z_i \in [0,1]$,
where 
$Z_i \in \{0,1\}$ is the allele value of the $i$th noise individual.
This is motivated by the fact that, if the lab uses shotgun sequencing technologies \cite{nextgen}, each reading of site $s$ is effectively a $\Ber\left(\alpha_0 X + \sum_{i=1}^K \alpha_i Z_i\right)$ random variable.
Via repeated readings of $s$, $Y$ can thus be obtained with arbitrary accuracy, so we assume for simplicity that $Y$ is observed exactly.
\bluechange{Upon receiving $Y$ from the lab, since Alice knows the mixing coefficients $\alpha_0,\dots,\alpha_K$ and the noise genotypes $Z_1,\dots,Z_K$, Alice can subtract the noise term from $Y$ and divide by $\alpha_0$ to recover $X$.
Alternatively, if the ``noise  individuals''  are not willing to share their genotypes $Z_i$ with Alice, 
a secure multiparty computation scheme \cite{Yao1986, Goldreich1987, Lindell2020} 
can be used to 
compute the noise $\sum_{i=1}^K \alpha_i Z_i$ only, and Alice can use the result to remove the noise contribution from $Y$.
}
We refer to Section~\ref{sec:discussion} for additional motivation and discussion on model assumptions.
\bluechange{We point out that the noise individuals can also be understood as creating an effective (discrete) channel from $X$ to the observation $Y$. 
As illustrated in Figure \ref{fig:Overview}(b), different choices of mixing coefficients lead to different channels.}


\longversion{
We refer to the long version of this manuscript \cite{privateDNAlong} for a more detailed discussion on this model.
\kmcomment{I guess here we can give more detain about the model}
}




Following \cite{Maddah-Ali}, we model $X,Z_1,\dots,Z_K$ as i.i.d.~random 
variables with $\Pr(X=1) = p \in [0,0.5]$.
We refer to $p$ as the minor allele frequency, a parameter that is known in practice for genetic loci of interest.
As in \cite{Maddah-Ali}, 
we utilize the mutual information as our privacy metric.
If we let $\alpha = (\alpha_0,...,\alpha_K)$ and $Z_\alpha = \sum_{k=1}^K \alpha_k Z_k$, our goal is thus to solve
\al{
\inf_{\alpha \in \R^{K+1} : \; \alpha > 0, \; \sum_{i=0}^K \alpha_i = 1} 
I\left( X; \alpha_0X+Z_\alpha\right),
\label{eq:main}
}
i.e., choose the mixing coefficients $\alpha_0,...,\alpha_K$ to minimize the mutual information between $X$ and the lab's observation. 
\bluechange{Here, $\alpha>0 $ means that all entries of $\alpha$ are strictly positive.
We point out that \eqref{eq:main} does not tend to zero by letting $\alpha_0 \to 0$, due to the discrete nature of the observation $Y$ (see Appendix~\ref{sec:notzero} for a simple proof of this fact). 
In addition, we note that this formulation assumes that the mixing coefficients can be chosen with arbitrary precision, which is not the case in practice.
We refer to Section~\ref{sec:discussion} for additional discussion on this point.}
We also further discuss the connection between our problem formulation and the work in \cite{Maddah-Ali, Maddah-Ali2} in more detail in Section~\ref{sec:related}.


The problem in (\ref{eq:main}) is equivalent to maximizing the conditional entropy $H(X|Y)$; i.e., the residual uncertainty in $X$ after observing $Y = \alpha_0 X+Z_\alpha$, and can thus be understood as maximizing the privacy of $X$.
It can also be thought of as the problem of finding an optimal noise $Z_\alpha$ among those of the form $\sum_k \alpha_k Z_k$, with $Z_k$ being i.i.d.~$\Ber(p)$. 
\bluechange{This kind of problem is usually  referred to as finding the worst-case noise \cite{Shamai1992, Verdu1996, Diggavi2001, Shomorony2013} 
but in our problem, this can be thought of as finding the best-case noise to maximize Alice's privacy. } 
One of our main results is that the solution to (\ref{eq:main}) is 
lower-bounded as
\al{
\inf_{\alpha \in \R^{K+1} : \; \alpha > 0, \; \sum_{i=0}^K \alpha_i = 1} I\left(X;\alpha_0 X + Z_\alpha\right)
\geq I\left(X;X + G\right),
\label{eq:lower}
}
where $G \sim \Geom\left( (1-p)^K \right)$ and $G$ is independent of $X$.
The right-hand side of (\ref{eq:lower}) can be computed explicitly as a function of $p$ and $K$.
Moreover, we verify empirically that this lower bound is very close to an upper bound provided by a greedy algorithm that  sets $\alpha_0 = 1$ and selects $\alpha_1,\alpha_2,...,\alpha_K$ sequentially to minimize the resulting mutual information, establishing $I\left(X;X + G\right)$ as a good approximation to the solution of (\ref{eq:main}).
We derive (\ref{eq:lower}) via a convex relaxation of (\ref{eq:main}), and 
use KKT conditions to show the lower bound.
In Section \ref{sec:efficient}, we derive a more computationally efficient algorithm to compute a weaker upper bound that also captures the general shape of the lower bound.

\section{Problem Setting and Preliminaries}
\label{sec:preliminaries}

Our goal is to characterize the mixing proportions $\alpha_0,\dots,\alpha_K$ that minimize the mutual information in (\ref{eq:main}).
Notice that we 
do not need to constrain the mixing proportions to add up to $1$, since scaling the observation $Y$
does not change the mutual information.
As a result, we can restrict ourselves to solving the optimization problem
\al{
\inf_{\alpha \in \R^{K+1} : \; \alpha > 0} I(X;\alpha_0 X+Z_\alpha),
\label{eq:main2}
}
where $\alpha = (\alpha_0,\dots,\alpha_K)$, $Z_\alpha = \sum_{i=1}^K \alpha_i Z_i$, and $X,Z_1,\dots,Z_K$ are independent ${\rm Ber}(p)$ random variables.

The optimization problem in (\ref{eq:main2}) 
is surprisingly complex. 
\bluechange{The fact that $\alpha_0 > 0$ can be chosen arbitrarily small may suggest that this mutual information can be driven to zero by letting $\alpha_0 \to 0$, but that is not the case (see Appendix~\ref{sec:notzero} for a simple proof of this).}
The symmetry between the variables $\alpha_0,\dots,\alpha_K$ may  suggest that $\alpha_i = 1$ for $i=0,\dots,K$  
would be an optimal solution.
However, a brute-force solution to (\ref{eq:main2}) over integer $\alpha_i$s for small values of $K$ shows that optimal solutions $(\alpha_0,\dots,\alpha_K)$ vary widely for different values of $p$,
as illustrated in Figure \ref{fig:optimal_scheme_K5_minor_allele}.  
Observe that the curve appears to be only piecewise smooth. At $p = 0.5,$ the optimal solution is given by $[1,1,2,4,8,16],$ at $p=0.25,$ the optimal solution is given by $[1,1,1,2,3,4],$ and at $p = 0.01,$ the optimal solution is given by $[1,1,1,1,1,1]$.  

As it turns out, the optimal solution to (\ref{eq:main2}) can be exactly characterized in the two extreme cases of $p$.
More precisely, if we define the \emph{uniform} scheme to be 
$\alpha_i = 1$ for $i=0,\dots,K$,
and we define the \emph{binary} scheme to be $\alpha_0 = 1$ and $\alpha_i = 2^{i-1}$ for $i=1,\dots,K$,
we have the following result.
\begin{theorem}\label{thm:extremes}
Fix some $K \in \N$.
Then there exists some $p^* > 0$ such that the uniform scheme is optimal for $p < p^*$.
Moreover, the binary scheme is optimal for $p=0.5$.
\end{theorem}
The two statements in Theorem~\ref{thm:extremes} are divided into Lemma \ref{lem:optimality_of_binary} and Lemma \ref{lem:optimality_of_uniform}, which are proved in Appendix \ref{sec:lemmas}.
Aside from the cases $p = 0.5$ and $p \approx 0$, there does not appear to be a simple expression for the optimal solution $\alpha$.

\bluechange{
Intuitively, the choice of optimal schemes for the extreme cases in Theorem~\ref{thm:extremes} can be justified as follows.
For fixed $K$ and $p$ small enough, it is most likely that all variables $X, Z_1, Z_2, \dots, Z_K$ will take value zero. The next most likely situation is that exactly one of these variables takes value one.  To create the most uncertainty for small $p$, it intuitively makes sense to set $\alpha_i = 1$ for all $i$ (i.e. the uniform scheme) because in this case, when any one (and exactly one) of $X, Z_1, Z_2, \dots, Z_K$ equals one,  $X$ could have been either zero or one from the sequencer's perspective based on the observation $\alpha_0 X + \sum_{i=1}^K \alpha_i Z_i$.  On the other hand, if $\alpha_i \neq \alpha_0$ and $Z_i$ is the only random variable that takes value one, then the sequencer knows that $X=0$.

When $p = 0.5$, we have that the distribution of $\sum_{i=1}^K \alpha_i Z_i$ is uniform on its support when $\alpha_i = 2^{i-1}$ (i.e. the binary scheme).  Therefore, the probability of the observation $\alpha_0 X + \sum_{i=1}^K \alpha_i Z_i$ taking value in  $j \in \{1, ..., 2^K-1\}$ is \begin{align} 
\Pr \left( \sum_{i=1}^K \alpha_i Z_i = j \right) \Pr(X = 0) + \Pr \left( \sum_{i=1}^K \alpha_i Z_i = j-1 \right) \Pr(X = 1).
\end{align}
When the sequencer observes a value of $\alpha_0 X + \sum_{i=1}^K \alpha_i Z_i$ in the range $j \in \{1, \dots, 2^K-1\}$, it is equally likely that $X$ was zero or one since $\Pr \left( \sum_{i=1}^K \alpha_i Z_i = j \right) \Pr(X = 0) = \Pr \left( \sum_{i=1}^K \alpha_i Z_i = j-1 \right) \Pr(X = 1)$.  Therefore, the uncertainty in $X$ is maximized for the binary scheme when $p = 0.5$.
}

\bluechange{It should be noted that increasing $K$ always decreases the optimal value of (\ref{eq:main2}).  This is because for any choice of positive choice of $\alpha_0, \dots, \alpha_{K-1}$, adding a $K$th noise individual with  $\alpha_K = \sum_{i=0}^{K-1} \alpha_i$ strictly decreases the mutual information in the objective of (\ref{eq:main2}).  Interestingly, however, if 
a constraint of the form $\alpha_0 \geq \kappa$ is imposed for some $\kappa > 0$,
then increasing $K$ (and forcing the new coefficient $\alpha_K$ to be positive) can increase the optimal value as discussed in Section~\ref{sec:utility_constraint}.
}


\vspace{1mm}
\noindent
\textbf{Notation:} Throughout the paper we use $\N$ to denote the set of natural numbers excluding $0,$ $\N_0$ to denote the set of natural numbers including $0,$ and $[N]$ to denote the set of natural numbers in $\{1, 2, \dots,  N\}$ for an integer $N.$
For a vector $v,$ $v > 0 $ means that all entries of $v$ are positive.
\longversion{
\iscomment{
To discuss if we have space: convolution of discrete measures, exponential complexity of brute-force search,
}
}

\longversion{

Notice that Equation \ref{eq:main} behaves very differently from the case where $Z_i$ is Gaussian.  In the Gaussian case, each $\alpha_i$ would be chosen as large as possible for all $i \in [K].$  However in our problem, if $\alpha_i > 1$ for all $i \in [K],$ then   
\begin{align} 
& I(X \; ; \; X + \sum_{i=1}^{K} \alpha_i Z_i) 
\nonumber \\ & = H(X) - H(X | X + \sum_{i=1}^{K} \alpha_i Z_i) = H(X) 
\end{align}
i.e. the scheme does not hide $X$ at all.  Therefore, if $\alpha_0 = 1,$ the solution to Equation \ref{eq:main} must have $\alpha_i = 1$ for at least one $i \in [K].$ 

\rdcomment{i might advocate for a little more discussion here: perhaps talk about how convolution of measures for discrete distributions is fundamentally a different problem than convolutions of continuous measures for this reason. i'll defer to IS's opinion on this tho.}

}

\begin{figure}[t]
\centering

\tikzstyle{every pin}=[fill=white,
	draw=black,
	font=\footnotesize]
	
\begin{tikzpicture}[scale=0.8]
\begin{axis} 
[width=0.46\textwidth,
height=0.39\textwidth,
legend cell align={left}, 
ylabel = mutual information, 
xlabel = minor allele frequency ($p$),
ymax = .28,
yticklabel style={
  /pgf/number format/precision=3,
  /pgf/number format/fixed},
]

\addplot [
line width=1.5pt,
dashed,
color=red!80!black
] table[x index=0,y index=1] {data/plt1_uniform.dat};
\addlegendentry{uniform}

\addplot [
line width=1.5pt,
dashed,
color=green!80!black
] table[x index=0,y index=1] {data/plt1_binary.dat};
\addlegendentry{binary}

\addplot [
line width=1.5pt,
color=blue!80!black
] table[x index=0,y index=1] {data/plt1_optimal.dat};
\addlegendentry{optimal}

\node[coordinate,pin=right:{$[1,1,1,1,1,1]$}] 
		at (axis cs:.01, 0.0423605362433) {};

\node[coordinate,pin=5:{$[1,1,1,2,3,4]$}] 
		at (axis cs:.25, 0.0942663435378) {};

\node[coordinate,pin=192:{$[1,1,2,4,8,16]$}] 
		at (axis cs:.5, 0.03125) {};

\end{axis}   

\end{tikzpicture}
\caption{
Optimal value of 
(\ref{eq:main2}) for integral $\alpha_i$s compared to the uniform scheme and the binary scheme,
for $K=5$ and $p \in [0,0.5]$.
At $p = 0.5,$ the optimal scheme is $\alpha = [1,1,2,4,8,16].$  At $ p = 0.25,$ the optimal scheme is $\alpha = [1,1,1,2,3,4].$  At $p = 0.01,$ the optimal scheme is $\alpha = [1,1,1,1,1,1].$
}
\label{fig:optimal_scheme_K5_minor_allele}
\end{figure}


\section{Main Results}


In order to tackle the discrete optimization problem in (\ref{eq:main2}) in the entire interval $p \in [0,0.5]$, we seek to bound its optimal solution.
Our main lower bound result is the following:


\begin{theorem} \label{thm:main_lower_bound}
	For any $K\in \mathbb{N}$ and $p \in [0, 0.5],$  we have 
	\begin{align} 
	& \inf_{\alpha \in \R^{K+1} \; : \; \alpha > 0}
	I(X \; ; \; \alpha_0 X +Z_\alpha) 
	\geq 
	I(X \; ; \; X + G),
	\label{eq:mainthm}
	\end{align}
	where $G$ is a geometric random variable, independent of $X$, with
	$\Pr(G = i) = (1-p)^K(1 - (1-p)^K)^i$ for $i \in \N_0$.
\end{theorem}

\bluechange{We note that the lower bound in \eqref{eq:mainthm} seems related in flavor to the well-known fact that the geometric distribution is the entropy-maximizing integer-valued distribution given a mean constraint.
However, notice that $Z_\alpha$ is not restricted to the integers, has no mean constraint, and its support changes as a function of $\alpha$.
Hence, the proof of Theorem~\ref{thm:main_lower_bound} is much more sophisticated than standard worst-case noise characterizations via entropy maximization arguments~\cite{coverthomas}.
At a high level, we prove the lower bound
in Theorem \ref{thm:main_lower_bound} by forming a convex relaxation of the minimization problem, perturbing the relaxation to form a problem that is analytically solvable using KKT conditions, and using perturbation analysis to find a lower bound on the relaxation.  
}

Intuitively, Theorem~\ref{thm:main_lower_bound} says that the noise distribution of $G$ is more effective at hiding $X$ than the optimal noise $Z_\alpha$.
This lower bound can in fact be explicitly computed as
\begin{align}
& I(X \; ; \; X + G)
= H(p) 
\nonumber \\ & - (p-1) \left( (1-p)^K - 1\right) \log \left( \frac{1-(1-p)^{K+1} }{(p-1) \left( (1-p)^K -1 \right)}\right) 
\nonumber \\ & - p \log \left( \frac{1-(1-p)^{K+1}}{p} \right).
\label{eq:closedform}
\end{align}
Observe that this formula is quickly computable for any value of $K$ and $p,$ making it attractive from a computational standpoint.  To assess how tight the lower bound is, we empirically compare it to an upper bound that is computed with a greedy algorithm in Figure \ref{fig:greedy_versus_bound_K15}.  For a given $p$ and $K,$ the greedy algorithm chooses $\alpha_0 = 1, \; \alpha_1 = 1,$ and sequentially chooses \begin{align} \alpha_{j} = \argmin_{a \in \mathbb{N} \; : \; 1 \leq a \leq 1 + \sum_{i = 1}^{j-1}  \alpha_i} I\left(X \; ; \; X + a Z_{j} + \sum_{i = 1}^{j-1} \alpha_i Z_i\right)
\end{align}
for $2 \leq j \leq K$.  At the $j$th step we consider all values of $\alpha_{j}$ between $1$ and $1 + \sum_{i = 1}^{j-1} \alpha_i$ because setting $\alpha_j > 1 + \sum_{i = 1}^{j-1} \alpha_i$ can not decrease the mutual information from when $\alpha_j = 1 + \sum_{i = 1}^{j-1} \alpha_i.$  
\bluechange{This is straightforward to verify, and the proof of Lemma \ref{lem:binary_scheme_performance} illuminates why this is the case.}


As seen in Figure \ref{fig:greedy_versus_bound_K15}, $I(X \; ; \; X + G)$ serves as a tight lower bound when compared with the upper bound.  
A similar picture can be obtained for other values of $K$.
\longversion{
Therefore, we can think of $I(X \; ; \; X - G)$ and its closed-form expression (\ref{eq:closedform}) as an approximation to the solution of (\ref{eq:main2}).  }
This is surprising, because for a given $K$, it is not possible in general to choose $\alpha_i$s to make the pmf of $\sum_{i = 1}^{K} \alpha_i Z_i$ look like 
the pmf of  $G$ (or a shifted version of it), as illustrated in Figure~\ref{fig:greedy_versus_geom_pmf_K15}.

\begin{figure}[t]
\centering

\begin{tikzpicture}[scale=0.8]
\begin{axis} 
[width=0.47\textwidth,
height=0.36\textwidth,
legend cell align={left}, 
ylabel = mutual information, 
xlabel = minor allele frequency (p),
ymax = .06]

\addplot [
line width=1.5pt,
color=red!80!black
] table[x index=0,y index=1] {data/plt2_greedy.dat};
\addlegendentry{upper bound (greedy)}

\addplot [
line width=1.5pt,
color=blue!80!black,
dashed
] table[x index=0,y index=1] {data/plt2_bound.dat};
\addlegendentry{lower bound}

\end{axis}    
\end{tikzpicture}
\caption{Comparison between the lower bound from (\ref{eq:mainthm}) and the upper bound provided by the greedy algorithm for $K = 15$.}
\label{fig:greedy_versus_bound_K15}
\end{figure}

\begin{figure}[h]
\centering

\begin{tikzpicture}[scale=0.8]
\begin{axis} 
[width=0.47\textwidth,
height=0.36\textwidth,
legend cell align={left}, 
legend pos = north east,
ymax = .05]


\addplot +[only marks,
		point meta=explicit symbolic,
		color=black!80!black]
table[x index=0,y index=1] {data/plt5_greedy_pmf_K15.dat};

\addlegendentry{greedy pmf}

\addplot +[only marks,
		point meta=explicit symbolic,
	color=black!80!black]
table[x index=0,y index=1] {data/plt5_geom_pmf_K15.dat};
\addlegendentry{geometric pmf}

\end{axis}    
\end{tikzpicture}

\caption{
Comparison of the pmf of $Z_\alpha$ produced by the greedy algorithm ($\alpha = [1, 1, 1, 1, 2, 2, 3, 4, 5, 6, 7, 8, 10, 12, 16, 19]$) and the (truncated) Geometric pmf in the lower bound (\ref{eq:mainthm}), for $K=15$ and $p=0.25$.
}
\label{fig:greedy_versus_geom_pmf_K15}
\end{figure}


While finding the greedy solution $(\alpha_0,\dots,\alpha_K)$ requires $\Omega(2^K)$ time in the worst case, similar plots to Figure \ref{fig:greedy_versus_bound_K15} can be obtained for larger values of $K$ using a more computationally efficient variation of the greedy algorithm.  
In Section \ref{sec:efficient}, we present a scheme that can be computed in $\text{poly}(K)$ time whose performance curve captures the general shape of the lower bound.  The recursion we prove to derive the algorithm is quite general, and could be of independent interest.



\longversion{

\begin{figure}[h]
\centering

\begin{tikzpicture}[scale=0.8]
\begin{axis} 
[width=0.47\textwidth,
height=0.36\textwidth,
title ={Ratio: Upper Bound to Lower Bound},
ylabel = mutual information, 
xlabel = minor allele frequency (p),
legend pos = north west]

\addplot [
line width=1pt,
color=red!80!black
] table[x index=0,y index=1] {data/plt4_greedy_opt_ratio_K15.dat};
\addlegendentry{K = 15}

\addplot [
line width=1pt,
color=blue!80!black
] table[x index=0,y index=1] {data/plt4_greedy_opt_ratio_K10.dat};
\addlegendentry{K = 10}

\addplot [
line width=1pt,
color=green!80!black
] table[x index=0,y index=1] {data/plt4_greedy_opt_ratio_K5.dat};
\addlegendentry{K = 5}

\end{axis}    
\end{tikzpicture}

\caption{}
\label{fig:greedy_versus_bound_K15}
\end{figure}

}












\section{Connection to Prior Work}
\label{sec:related}


\bluechange{
Mutual information has been used as a measure of information leakage as early as the 1970s in the study of the wiretap channel \cite{Csiszar1978, Wyner1975, Maurer1994, Maurer2001}, which has applications in wireless communications. It has found many other applications such as the analysis of  anonymous remailer protocols \cite{Chatzikokolakis2010} and database security \cite{Sankar2013}. 
Mutual information as a measure of information leakage has even been connected to other widely studied measures of privacy such as semantic security \cite{Bellare2012} and differential privacy \cite{Cuff2016, Kairouz2016}.  
}

Our problem formulation (\ref{eq:main}) is motivated by the problem studied in \cite{Maddah-Ali}.  In \cite{Maddah-Ali}, the authors consider the same high-level problem of providing privacy to genotype information through the mixture of distinct samples prior to sequencing.
But there are several differences in the focus of the analysis in \cite{Maddah-Ali}.
In addition to considering the privacy of 
Alice's genotype information, they also \bluechange{explicitly} consider the probability of Alice being able to correctly recover her genotype \bluechange{in the presence of sequencer noise.
Sequencer noise is included in the model by assuming that each reading the sequencer makes is incorrect with some probability.}
The authors of \cite{Maddah-Ali} propose a scheme that uses $U \in \mathbb{N}$ non-communicating sequencers to sequence $U$ unknown DNA samples with privacy and reconstruction guarantees.  Similar to our problem formulation, they also use $K$ noise individuals to generate privacy.  However, they do not study the problem of optimizing the proportions of each DNA sample sent to the sequencers to maximize privacy.  
\bluechange{In contrast, we focus on understanding how the proportions of samples in the mixture sent to a sequencer can be chosen to optimize the privacy of a single allele in a particular unknown genome, without explicitly considering sequencer noise or the reconstruction condition.  While we focus on optimizing privacy from a single sequencer receiving a mixture of samples, our analysis is applicable to the case where there are multiple non-communicating sequencers as in \cite{Maddah-Ali} since in this case, the privacy from each sequencer can be analyzed independently.}  
\bluechange{In Section \ref{sec:discussion}, 
we propose a utility constraint that takes reconstructability into account, and discuss how our results apply when this constraint is present.}  

The proposed solution in \cite{Maddah-Ali} involves the sequencing of the unknown DNA of $U$ individuals simultaneously using $U$ non-communicating sequencing laboratories and $K$ noise individuals. 
The mixture of the DNA samples of all $K$ noise individuals and all unknown DNA samples except the $i$th one is sent to the $i$th sequencing laboratory.  For each mixture sent to a sequencing laboratory, the included DNA samples are mixed in equal proportion.   
Observe that each sequencing laboratory observes a mixture that includes $K + U-1$ DNA samples.  
\bluechange{This scheme can also be analyzed in the framework of our paper. Observe that for a particular unknown DNA sample $A$ in a mixture sent to a particular sequencer in their scheme, all other samples (unknown and known) in the mixture act as noise in hiding $A$ from this sequencer. Thus, we can use our framework to analyze the privacy of $A$ from this sequencer by treating all other samples in the mixture as noise individuals.  Interestingly, the scheme used in \cite{Maddah-Ali} is the uniform scheme in the language of our paper, which we showed generates optimal privacy for $p$ close to $0$.}



In \cite{Maddah-Ali2}, 
the unknown DNA samples of $U$ individuals 
are mixed  with the samples of $K=U$ noise individuals and then sequenced using one sequencing laboratory. 
Both the $i$th unknown sample and the $i$th noise sample are mixed in with amount $\alpha_i = 2^i $ 
for $i = 0,1, ..., K-1$. 
\bluechange{It is interesting that this is similar to the binary scheme, which we showed is optimal in our problem formulation for $p = 0.5.$}

\section{Proof of Theorem~\ref{thm:main_lower_bound}}




We obtain the lower bound on $I(X \, ; \, \alpha_0 X + Z_{\alpha})= H(p) - H(X \, | \, \alpha_0 X+Z_{\alpha})$ 
by finding a lower bound on $-H(X \, | \, \alpha_0 X+Z_{\alpha})$.  
Therefore, we consider
\begin{align}
   \label{eq:integral_alpha_opt_prob}
    & \inf_{\alpha \in \R^{K+1}  : \: \alpha >0} -H\left(X \; \left| \; \alpha_{0} X+\sum_{k=1}^K \alpha_k Z_k \right. \right).
\end{align}
Observe that the pmf of the random variable $Z_{\alpha} = \sum_{i = 1}^K \alpha_i Z_i$ has probability $(1-p)^{K}$ at its lowest support value. 
More precisely, $0$ is the minimum value that $Z_\alpha$ can take, which occurs with probability $(1-p)^{K}$.
A relaxation to (\ref{eq:integral_alpha_opt_prob}) can then be obtained by ignoring all constraints on the pmf of $Z_\alpha$ except the constraint on the minimum pmf value.
Thus, for a fixed value of $\alpha_0$, a relaxation to (\ref{eq:integral_alpha_opt_prob}) is given by
\begin{align}
   \label{eq:relax_inf1}
    \inf_{Q} \; &  -H\left(X \; | \; \alpha_0 X+Q\right)
    \\\text{subject to:} \; & 
    \nonumber 
    Q \text{ is a discrete random variable} \nonumber
    \\ & \text{$Q$ is independent of $X$} \nonumber
    \\ & \Pr(Q = 0) = (1-p)^K \nonumber
    \\ & \Pr(Q = i) = 0 \; \text{ for $i < 0$}.
    \nonumber
\end{align}
Furthermore, as we prove in Lemma~\ref{lem:alpha0_condition} in Appendix~\ref{sec:lemmas}, 
fixing $\alpha_0 = 1$ in (\ref{eq:relax_inf1}) and constraining the support of $Q$ to be integer 
does not change the optimal value.  We assume these additional constraints throughout.
    

Let $q_{(i)}$ be the pmf of $Q$; i.e., $q_{(i)} = \Pr(Q=i)$ for $i \geq 0$.
In order to write (\ref{eq:relax_inf1}) explicitly in terms of $q$ under the assumption that $Q$ has integer support and $\alpha_0 = 1$, we define 
\begin{align}
    g_j(q) & \triangleq -H(X|X+Q=j)\Pr(X+Q=j) \nonumber \\
    & =  \Pr(Q+X=j)  \nonumber \\
    & \quad \times \bigg[ \frac{(1-p)q_{(j)}}{\Pr(Q+X=j)} \log\left(\frac{(1-p)q_{(j)}}{\Pr(Q+X=j)}\right) \nonumber \\       &  \quad + \frac{p q_{(j-1)}}{\Pr(Q+X=j)} \log\left(\frac{p q_{(j-1)}}{\Pr(Q+X=j)}\right) \bigg] \nonumber \\   
    & = (1-p)q_{(j)} \log \left(\frac{(1-p)q_{(j)}}{(1-p)q_{(j)} + p q_{(j-1)}} \right) 
    \nonumber 
    \\ & \quad + p q_{(j-1)} \log \left(\frac{p q_{(j-1)}}{(1-p)q_{(j)} + p q_{(j-1)}} \right).
\end{align}
Assuming integer support for $Q$ and fixing $\alpha_0 = 1,$ we have that (\ref{eq:relax_inf1}) written in terms of $g_j(q)$ is given by
\begin{align}
\label{eq:relax_inf2}
\inf_{q_{(i)} \geq 0 : \; i \in \N_0 } \; & \quad \sum_{j \in \mathbb{N}} g_j(q)
\\ \text{subject to:} \; & \quad q_{(0)} = (1-p)^K \nonumber \\ &  \quad    \sum_{j=0}^\infty q_{(j)} = 1. \nonumber
\end{align}  

Observe that (\ref{eq:relax_inf2}) is a convex minimization problem with infinitely many variables.  
For such problems, to the best of our knowledge, a solution to the KKT conditions is not in general guaranteed to yield an optimal solution.
For that reason, we do not seek to directly solve the KKT conditions and, instead, we consider a \emph{support-constrained} version of (\ref{eq:relax_inf2}), where the support of the pmf of $Q$ is restricted to $\{0,...,n\}$, and let $n \to \infty$.
%
%
The support-constrained version of (\ref{eq:relax_inf2}) is given by 
\begin{align}
\label{eq:relax_const}
\inf_{q_{(i)} \geq 0 : \; i \in \{0,...,n+1\} } \; & 
\sum_{j =1}^{n+1} g_j(q) \\  \text{subject to:}  \; &  
\; q_{(0)} = (1-p)^K, \quad q_{(n+1)} = 0 \nonumber \\
\quad & \; \sum_{j=0}^{n+1} q_{(j)} = 1. \nonumber
\end{align}  
Note that this is no longer a relaxation to the original problem.

In order to ensure that the derivative of the objective function exists at all feasible $q_{(i)},$ and ultimately find a lower bound on the optimal value of (\ref{eq:relax_const}) through perturbation analysis, we change the constraints in (\ref{eq:relax_const}) from $q_{(i)} \geq 0$ to $q_{(i)} > 0$ and from $q_{(n+1)} = 0$ to $q_{(n+1)} = \epsilon$ where 
$\epsilon > 0$, obtaining
\begin{align}
\label{eq:relax_const_posq}
\inf_{q_{(i)} > 0 : \; i \in \{0,...,n+1\} } \; &
\sum_{j =1}^{n+1} g_j(q)
\\ \text{subject to:} \; & 
\; q_{(0)} = (1-p)^K, \quad q_{(n+1)} = \epsilon \nonumber \\ 
\quad  &\;   \sum_{j=0}^{n+1} q_{(j)} = 1. \nonumber
\end{align}  

Let $V_n^*$ be the optimal value of (\ref{eq:relax_const}) and let $V_{n,\epsilon}^*$ be the optimal value of (\ref{eq:relax_const_posq}) for a given $\epsilon.$ Due to the continuity of the objective function in (\ref{eq:relax_const}), we have $ V_n^* \geq \inf_{\epsilon >0} V_{n,\epsilon}^*.$  More precisely, for any solution to  (\ref{eq:relax_const}), it follows that (\ref{eq:relax_const_posq}) can get arbitrarily close to the corresponding value as $\epsilon \to 0$ due to the continuity of the objective function.

While we do not know the solution to (\ref{eq:relax_const}) or (\ref{eq:relax_const_posq}), we will perturb (\ref{eq:relax_const_posq}) to form a problem we can solve analytically, then use perturbation analysis to lower bound the optimal value of (\ref{eq:relax_const_posq}), and ultimately lower bound the optimal value of (\ref{eq:relax_const}).  Let $\beta \triangleq (1-p)^K.$  The perturbed version of (\ref{eq:relax_const_posq}) is given by 
\begin{align}
\label{eq:relax_const_pert_posq}
\inf_{q_{(i)} > 0 : \; i \in \{0,...,n+1\} } & \;
\sum_{j =1}^{n+1} g_j(q) \\	
\text{subject to:} 
& \; \; q_{(0)} = \beta , \quad q_{(n+1)} = \beta  (1 - \beta )^{n+1}
\nonumber \\ &  \;  \sum_{j=0}^{n+1} q_{(j)} = 1 - (1 - \beta )^{n+2}. \nonumber
\end{align}  
Observe that as $n$ increases, we have that $\beta  (1 - \beta )^{n+1}\to 0$ and $1 - (1 - \beta )^{n+2} \to 1.$  In other words, the constraints in (\ref{eq:relax_const_pert_posq}) approach the constraints in (\ref{eq:relax_const}).
We can solve (\ref{eq:relax_const_pert_posq}) by solving the KKT conditions since the problem is convex, Slater's condition holds, and the objective function and constraint functions are differentiable on the domain $q_{(i)} > 0.$  
Let $f_0(q)$ be the objective function of (\ref{eq:relax_const_pert_posq}).  The Lagrangian is given by 
\begin{align}  \label{eq:lagrangian}
L(q, v, \lambda) = f_0(q) 
& + v_1 \left(\sum_i q_{(i)} - 1 + (1 - \beta)^{n+2} \right) 
\nonumber \\ & + v_2 \left(q_{(0)} - \beta \right) 
\nonumber \\ & + v_3 \left(q_{(n+1)} - \beta (1 - \beta)^{n+1} \right). 
\end{align}
\longversion{
\iscomment{what is $\lambda$ here?}
}
The perturbation values in (\ref{eq:relax_const_pert_posq}) are carefully chosen so that the KKT conditions yield an optimal solution given by 
\begin{align}
\label{eq:KKT_conditions_sol}
& q_{(i)} =  \beta (1 - \beta)^{i}  \text{ for } i \in \{0,...,n+1\}.
\end{align}
This corresponds to the first $n+2$ terms of the pmf of a Geometric distribution.
The derivation of (\ref{eq:KKT_conditions_sol}) and the optimal Lagrange multipliers $v_1^*,v_2^*,v_3^*$ are provided in Lemma~\ref{lem:solve_KKT}.

Let ${U}_{n}^*$ be the solution to (\ref{eq:relax_const_pert_posq}), obtained by plugging in (\ref{eq:KKT_conditions_sol}).
%
\longversion{
Applying the geometric sum formula and simplifying,  the optimal value for (\ref{eq:relax_const_pert_posq}) is   
\begin{align} 
& -(1-p) \beta (1-\beta) \frac{1 - (1-\beta)^{n+1}}{1 - (1-\beta)} \log \left( \frac{(1-p) (1-\beta) + p }{(1-p) (1-\beta)} \right) 
\nonumber \\ & - p \beta \frac{1 - (1-\beta)^{n+1}}{1 - (1-\beta)} \log \left( \frac{(1-p)  (1-\beta) + p }{ p } \right).
\end{align}
}
Using the perturbation analysis from Section 5.6.1 of \cite{boyd}, we see that the optimal value of (\ref{eq:relax_const_posq}), $V_{n,\epsilon}^*$, is lower bounded as
\begin{align}
\label{eq:relax_const_posq_bound}
V_{n,\epsilon}^* \geq
U_n^*
- v_1^* \left( (1 - \beta)^{n+2} \right) - v_3^* \left(- \beta (1 - \beta)^{n+1} +\epsilon \right),
\end{align} 
where $v_1^*$ and $v_3^*$ are the optimal Lagrange multipliers for (\ref{eq:relax_const_pert_posq})
described in Lemma~\ref{lem:solve_KKT} in Appendix~\ref{sec:lemmas}.
Taking the infimum of (\ref{eq:relax_const_posq_bound}) over 
$\epsilon > 0$ then
yields 
\begin{align}
\label{eq:relax_const_bound}
V_n^* \geq
U_n^*
- v_1^* \left( (1 - \beta)^{n+2} \right) - v_3^* \left(- \beta (1 - \beta)^{n+1} \right).
\end{align} 

The sequence $V_n^*$ of optimal values returned by (\ref{eq:relax_const}) is non-increasing in $n$.  
Thus, letting $n \to \infty$ in (\ref{eq:relax_const_bound}) implies that $\lim_{n\to\infty}U_{n}^*$ is a lower bound to $-H(X \, | \, \alpha_0 X + Z_{\alpha})$ for any choice of $\alpha_i$'s.
Notice that, as $n \to \infty$, $q_{(j)}$ in (\ref{eq:KKT_conditions_sol}) converges to the pmf of a Geometric random variable $G$ with 
$\Pr(G = i) = \beta (1-\beta)^i = (1-p)^K(1 - (1-p)^K)^i$ for $i \in  \N_0$.
Since the objective function of (\ref{eq:relax_const_pert_posq}) is $-H(X|X+Q)$ where $Q$ has pmf $q_{(j)}$, this concludes our proof.


\longversion{
Thus, taking the limit of the lower bound (\ref{eq:relax_const_bound}) as  $n \to \infty$ and simplifying yields the following lower bound on $-H(X \; | \; \alpha_0 X + Z_{\alpha})$ for any choice of $\alpha_i$'s: \begin{align} 
\label{eq:lower_bound_neg_cond_ent}
& -(p-1) \left( (1-p)^K - 1\right) \log \left( \frac{1 - (1-p)^{K+1} }{(p-1) \left( (1-p)^K -1 \right)}\right) 
\nonumber \\ & - p \log \left( \frac{1 - (1-p)^{K+1} }{p} \right).
\end{align}

Finally, observe that
$-H(X \; ; \; X - G)$ is equal to (\ref{eq:lower_bound_neg_cond_ent})
where $G$ is a geometric random variable with
$\Pr(G = i) = (1-p)^K(1 - (1-p)^K)^i$ for $i \in  \N_0.$
}


\longversion{
\section{Relationship with worst-case noise}

When is geometric worst-case noise from capacity standpoint?
For a fixed channel standpoint?
}

\section{Efficiently Computable Scheme}
\label{sec:efficient}

Because the uniform scheme is optimal for $p \to 0$ and the binary scheme is optimal for $p = 0.5,$  it is natural to try and combine these schemes to interpolate the performance for $p$ in the range $0<p<0.5.$  Combining only the uniform scheme and the binary scheme does not produce a scheme with performance similar to the lower bound.  However, the general shape of the lower bound can be achieved by a polynomial-time computable scheme that combines the uniform scheme, the linear scheme, and the binary scheme. The binary scheme and uniform scheme are defined in Section \ref{sec:preliminaries}, and the linear scheme is defined by setting $\alpha_0 = 1$ and $\alpha_i = i$ for $i \in [K].$  We define the $(K,L,U)$-binary-linear-uniform scheme as $\alpha_0 = 1$ and 
\begin{align} &\alpha_i = 1 \quad i \in \{1, ..., U\} \nonumber \\  &\alpha_{U+i} = i \quad i \in \{1, ..., L\} \nonumber \\
&\alpha_{U+L+i} = 2^{i-1} \quad i \in \{1, ..., K-U-L\}.
\end{align}


For a given $K$ and $p,$ we define the $K$-binary-linear-uniform scheme as the $(K,L^*,U^*)$-binary-linear-uniform scheme where
\begin{align} & (L^*, U^*) = \argmin_{(L,U) \in ([K]\cup\{0\})^2 \; : \; L + U \leq K} \label{eq:blu_scheme} \\ &  I\left(X \; ; \; X + \sum_{i=1}^{U} Z_i + \sum_{j=1}^{L} j Z_{U+j} +  \sum_{k=1}^{K-U-L} 2^{k -1} Z_{U+L+k}\right) \nonumber.
\end{align}
The $K$-binary-linear-uniform scheme is computed in $O(K^5)$ time using a generalization of the recursion used to prove Lemma~\ref{lem:binary_scheme_performance}.  

\begin{figure}[h]
\centering
\begin{tikzpicture}[scale=0.8]
\begin{axis}
[width=0.47\textwidth,
height=0.36\textwidth,
title ={Comparison of Bounds: $K = 15$},
ylabel = mutual information, 
xlabel = minor allele frequency (p),
ymax = .6,
legend pos = north east]

\addplot [
line width=1pt,
color=red!80!black
] table[x index=0,y index=1] {data/plt2_greedy.dat};
\addlegendentry{greedy}

\addplot [
line width=1pt,
color=green!80!black
] table[x index=0,y index=1] {data/plt3_uniform.dat};
\addlegendentry{uniform}

\addplot [
line width=1pt,
color=purple!80!black
] table[x index=0,y index=1] {data/plt3_linear.dat};
\addlegendentry{linear}

\addplot [
line width=1pt,
color=orange!80!black
] table[x index=0,y index=1] {data/plt3_binary.dat};
\addlegendentry{binary}

\addplot [
line width=1pt,
color=yellow!80!black,
dashed
] table[x index=0,y index=1] {data/plt3_binary_linear_uniform.dat};
\addlegendentry{binary-linear-uniform}

\addplot [
line width=1pt,
color=blue!80!black,
dotted
] table[x index=0,y index=1] {data/plt2_bound.dat};
\addlegendentry{lower bound}

\end{axis}    
\end{tikzpicture}

\begin{tikzpicture}[scale=0.8]
\begin{axis}
[width=0.47\textwidth,
height=0.36\textwidth,
title ={Comparison of Bounds: $K = 15$},
ylabel = mutual information, 
xlabel = minor allele frequency (p),
ymax = .01]

\addplot [
line width=1pt,
color=red!80!black
] table[x index=0,y index=1] {data/plt2_greedy.dat};

\addplot [
line width=1pt,
color=green!80!black
] table[x index=0,y index=1] {data/plt3_uniform.dat};

\addplot [
line width=1pt,
color=purple!80!black
] table[x index=0,y index=1] {data/plt3_linear.dat};

\addplot [
line width=1pt,
color=orange!80!black
] table[x index=0,y index=1] {data/plt3_binary.dat};

\addplot [
line width=1pt,
color=yellow!80!black,
dashed
] table[x index=0,y index=1] {data/plt3_binary_linear_uniform.dat};

\addplot [
line width=1pt,
color=blue!80!black,
dotted
] table[x index=0,y index=1] {data/plt2_bound.dat};

\end{axis}    
\end{tikzpicture}
\caption{Comparison between the lower bound from (\ref{eq:mainthm}) and all upper bounds discussed in the paper for $K = 15$.  Observe that the upper bound corresponding to the $K$-binary-linear-uniform scheme captures the general shape of the lower bound.}
\label{fig:efficiently_computable_schemes}
\end{figure}

We compare the performance of the uniform scheme, the binary scheme, the $K$-binary-linear-uniform scheme, and the scheme generated by the greedy algorithm in Figure~\ref{fig:efficiently_computable_schemes}.  We observe that as $K$ increases, the scheme generated by the greedy algorithm is much tighter than the   $K$-binary-linear-uniform scheme after the initial hump in the lower bound.

To compute the $K$-binary-linear-uniform scheme, we compute $H(X \; | \; X + Z_\alpha)$ for each valid $Z_\alpha$ in the optimization in (\ref{eq:blu_scheme}) using a recursive algorithm.  To simplify the exposition in the description of the algorithm, we define $\tilde{X} = 2X - 1$,  $\tilde{Z}_i = 2Z_i - 1$, and $\tilde{Z}_\alpha = \sum_{i=1}^K \alpha_i \tilde{Z}_i$ so that all distributions in the derivation of the recursion are symmetric about the origin, and compute $H(\tilde{X} \; | \; \tilde{X} + \tilde{Z}_\alpha)$ since  $H(X \; | \; X + Z_\alpha) = H(\tilde{X} \; | \; \tilde{X} + \tilde{Z}_\alpha)$.  To describe this algorithm, we define
\begin{align*} 
    T_{M, N} = \tilde{X} + \sum_{i=1}^{M} \alpha_i  \tilde{Z}_i +  \sum_{j=1}^{N} 2^{j-1} \tilde{Z}_{M+j}
\end{align*}
where $M+N = K$ for some choice of $\alpha_i$ for $i \in [M]$,
and define
\begin{align*} 
    & A_M = \sum_{i=1}^{M} \alpha_i,
    \\ & 
    S_{M,N} = 1 + A_M + \sum_{j=1}^{N} 2^{j-1}.
\end{align*}
In Lemma \ref{lem:main_recursion}, we prove that for $M,N \in \mathbb{N}$ such that $M>0$ and $N>1,$ we have that 
	\begin{align} 
		& H(\tilde{X} \; | \; T_{M,N})  \nonumber
		\\ & = H(\tilde{X} \; | \; T_{M,N-1}) \nonumber
		\\ & - p \!\!\!\! \sum_{t=-S_{M,N-1}}^{-S_{M,N-1}+2A_M} \!\!\!\!\!\!\!\!\!\! \Pr(T_{M,N-1} = t) H(\tilde{X} \; | \; T_{M,N-1} = t) \nonumber
		\\ & - 	(1-p) \!\!\!\! \sum_{t=S_{M,N-1}-2A_M}^{S_{M,N-1}} \!\!\!\!\!\!\!\!\!\! \Pr(T_{M,N-1} = t) H(\tilde{X} \; | \; T_{M,N-1} = t) \nonumber
		\\ & + \sum_{t=-A_M}^{A_M} \Pr(T_{M,N} = t) H(\tilde{X} \; | \; T_{M,N} = t). \label{eq:main_rec}
\end{align}    
\begin{algorithm} \label{alg:uniform_binary_polytime}
	\SetAlgoLined
	\KwData{$N,$ $M,$ $p$}
	\KwResult{$I_{M,N}$}
	$n_1 \leftarrow \min(N, \ceil{\log(2A_M+2)})$\;
	$D_{n_1} \leftarrow$ array containing PMF for $\sum_{i=1}^M \alpha_i \tilde{Z}_i +  \sum_{j=1}^{n_1} 2^{j-1} \tilde{Z}_{M+j}$ starting at the lowest support value and ending at highest support value\;
	$D_{\text{low}} \leftarrow$ array containing first $2A_M+1$ entries of  $D_{n_1}$\;
	$D_{\text{high}} \leftarrow$ array containing last $2A_M+1$ entries of $D_{n_1}$\;
	$H_{n_1}  \leftarrow $ $\sum_{t=-S_{M,n_1}}^{S_{M,n_1}} \Pr(T_{M, n_1} = t) H(\tilde{X} \; | \; T_{M, n_1} = t)$ computed using $D_{n_1}$\; 
	$H_{\text{low}} \leftarrow $ 
	$\sum_{t=-S_{M,n_1}}^{-S_{M,n_1}+2A_M} \Pr(T_{M, n_1} = t) H(\tilde{X} \; | \; T_{M, n_1} = t)$ computed using $D_{\text{low}}$\; 
	$H_{\text{high}} \leftarrow $ $\sum_{t=S_{M,n_1}-2A_M}^{S_{M,n_1}} \Pr(T_{M, n_1} = t) H(\tilde{X} \; | \; T_{M, n_1} = t)$ computed using $D_{\text{high}}$\; 	
	$n \leftarrow n_1 + 1$\;
	\While{$n \leq N$}{
		$H_{n}  \leftarrow H_{n-1}$\;
		$H_{n} \leftarrow H_{n} - p H_{\text{low}} - (1-p) H_{\text{high}}$\;
		$D_{\text{low ext}} \leftarrow $ $D_{\text{low}}$ appended with $[0, \; 0]$\;
		$D_{\text{high ext}} \leftarrow $  $D_{\text{high}}$ prepended with $[0,\;  0]$\;
		$H_{\text{center}} \leftarrow$ 	$\sum_{t=-A_M}^{A_M} \Pr(T_{M, n} = t) H(X \; | \; T_{M, n} = t)$ computed using $(1-p) D_{\text{high ext}} + p D_{\text{low ext}}$\;
		$H_{n} \leftarrow H_{n} + H_{\text{center}}$\;
		$H_{\text{low}} \leftarrow (1-p) H_{\text{low}}$\; 
		$H_{\text{high}} \leftarrow  p H_{\text{high}}$\; 
		$D_{\text{low}} \leftarrow (1-p) D_{\text{low}}$\;
		$D_{\text{high}} \leftarrow p D_{\text{high}}$\;
		$n \leftarrow n + 1$\;
	}
	$I_{M,N} \leftarrow H(p) - H_N$

	\caption{Computation of $I(\tilde{X} \; ; \; T_{M,N})$}
\end{algorithm}
In order to apply the recursion in  (\ref{eq:main_rec}) 
to compute $H(\tilde{X} \; | \; T_{M, N})$, 
Algorithm \ref{alg:uniform_binary_polytime} first computes $H(\tilde{X} \; | \; T_{M, n})$ naively for $n = \min(N, \; \ceil{\log(2A_M+2)})$.  $H(\tilde{X} \; | \; T_{M, N})$ is then computed using (\ref{eq:main_rec})
along with the following properties of the summation terms.  

Observe that if $n >  \min(N, \ceil{\log(2A_M+2)}),$ then
\begin{align*}
& \sum_{t=-S_{M,n}}^{-S_{M,n}+2A_M} \Pr(T_{M, n} = t) H(\tilde{X} \; | \; T_{M, n} = t)
\\ & = p\sum_{t=-S_{M,n-1}}^{-S_{M,n-1}+2A_M} \Pr(T_{M, n-1} = t) H(\tilde{X} \; | \; T_{M, n-1} = t)
\end{align*}
and 
\begin{align*} 
& \sum_{t=S_{M,n}-2A_M}^{S_{M,n}} \Pr(T_{M, n} = t) H(\tilde{X} \; | \; T_{M, n} = t)
\\ & = 
(1-p) \sum_{t=S_{M,n-1}-2A_M}^{S_{M,n-1}} \Pr(T_{M, n-1} = t) H(\tilde{X} \; | \; T_{M, n-1} = t).
\end{align*}
Thus, both of these summations can be computed recursively once $n > \min(N,  \ceil{\log(2A_M+2)}).$

Finally, observe that 
\begin{align*} 	\sum_{t=-A_M}^{A_M} \Pr(T_{M, n} = t) H(\tilde{X} \; | \; T_{M, N} = t). 
\end{align*}
can be computed in $O(M^2)$ time and $O(M^2)$ space for each value of $n$ if the first $2A_M+1$ values of the PMF of  $ \sum_{j=1}^M \alpha_j \tilde{Z}_j +  \sum_{k=1}^{n-1} 2^{k-1} \tilde{Z}_{M+k}$ (starting with the smallest support value) and the last $2A_M+1$ values of the PMF of $ \sum_{j=1}^M \alpha_j \tilde{Z}_j +  \sum_{k=1}^{n-1} 2^{k-1} \tilde{Z}_{M+k}$ (ending with the largest support value)  are available.  

Algorithm \ref{alg:uniform_binary_polytime} performs the steps above $O(N)$ times. The algorithm must also build the PMF for $\sum_{j=1}^M \tilde{Z}_j$ which requires $O(M^3)$ time and $O(M^2)$ space.  Therefore, the algorithm runs in $O(N M^2 + M^3) = O(K^3)$ time and $O(M^2) = O(K^2)$ space.

There are $O(K^2)$ choices $(N, M)$ to check, so the solution to the optimization problem in (\ref{eq:blu_scheme}) can be computed in  $O(K^5)$ time and $O(K^2)$ space.

\section{Discussion} 
\label{sec:discussion}

In this paper, we studied how varying the proportions of individuals in the DNA mixture can affect the privacy guaranteed in the private DNA sequencing problem.  In specific, we derived a limit on the maximum possible privacy achievable, and compared this limit with the performance of several schemes obtained using algorithms we derived.
We ultimately observe that the performance curves generated by two of the algorithms are very close to the fundamental limit.

Next, we discuss several of the assumptions made in the problem formulation, their shortcomings, and how the model could be strengthened in future work.

\subsection{Assumptions on the Sequencing Model}
\label{sec:model_assumptions}

\begin{figure}[t]
\centering
	\includegraphics[width=\linewidth]{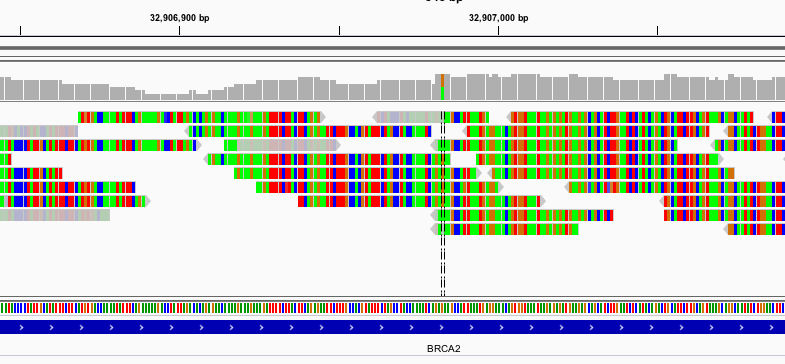}
	\caption{
Alignment of Illumina sequencing data from one individual to the BRCA2 gene (in chromosome 13) on the human reference genome, visualized with the Integrative Genome Viewer \cite{igv}.
In the dashed column ($\sim$32,906,980), out of the eight reads that cover it, four have the reference allele and four have the alternative allele.	
Given this data, one cannot infer whether it is the individual's father or mother chromosome that carries the minor allele.
	}
	\label{fig:seqdata}
\end{figure}

The model proposed in this paper represents an initial attempt at studying the problem of providing privacy to genetic information through the physical mixing of samples prior to sequencing.
While the idea of achieving privacy by forcing the sequencing lab to sequence a mixture of DNA samples may seem odd at first, to a certain extent, it already occurs in standard DNA sequencing pipelines.
A person's DNA is made up of two copies of each chromosome: a paternal chromosome and a maternal chromosome.
Hence, an individual's DNA can be thought of as a 50-50 mixture of the DNA of two unrelated individuals: the father and the mother.
Once a person's DNA is sequenced using next-generation sequencing technologies, the sequencing reads are aligned to the human reference genome, and genetic variants are identified.
These reads are equally likely to have come from the paternal and maternal chromosomes.
In Figure~\ref{fig:seqdata}, we show the alignment of Illumina sequencing data from one individual to the BRCA2 gene (in chromosome 13) \cite{brca2}.
In the highlighted position, out of the eight reads that cover it, four have the reference allele and four have the alternative allele.
This can be seen as a real-life illustration of the privacy strategy discussed in our paper, since it is not possible to know whether it is the person's father or mother that has the minor allele.
If the DNA samples of $K$ individuals were mixed, the picture will be similar to the one in Figure~\ref{fig:seqdata}, except that the minor allele counts will correspond to the mixture of $2K$ chromosomes.

In the model we study (and also the model in \cite{Maddah-Ali}), the allele at different loci are implicitly assumed to be independent.
In reality, genetic variants that are close in the genome are more likely to be inherited together, leading to what is known as \emph{linkage disequilibrium} \cite{Slatkin}.
This creates dependence across different alleles at different loci, which tends to reduce the privacy at any given locus.
One way to deal with this would be to extend 
our problem setup to simultaneously consider the privacy of a group of nearby correlated locations.
This is a natural direction for future work.

Our problem setup also relies on the fact that the locations that admit more than one allele are observed in separate reads.  
In the context of shotgun sequencing, this is equivalent to the locations being far enough apart in the genome so that a single read (150bp for standard Illumina platforms) cannot simultaneously cover two loci. 
This is often the case, as the number of single-nucleotide polymorphisms (SNPs) analyzed by standard genomic services is around one million, while the length of the human genome is roughly 3 billion.
Furthermore, standard direct-to-consumer genomic services utilize SNP arrays for sequencing, which essentially probe specific SNPs in the genome, rather than obtaining shotgun sequencing reads that could simultaneously cover multiple variants \cite{imai2011concordance}.



It should be noted that our current results do not take into account the noise from the sequencing machine itself.
Moreover, by assuming that the observation $Y$ is the precise proportion of the minor allele we are essentially assuming that a very high coverage is used for sequencing.
\bluechange{Of course, in practice, if $
\alpha_0$ is very small compared to $\alpha_i$s, it will not be detected by the sequencing machine.
To make the model more realistic, a noise term should be added to the observation $Y$.
Intuitively, once such a noise term is added, $\alpha_0$ will have to be selected to be larger than the noise level in order for it to be detectable at the output.
This will effectively impose an additional constraint on $\alpha_0$ in our optimization problem.
While we expect the ideas and techniques developed in this paper to generalize to that setting, we leave it as a direction for future work.
}

\subsection{Assumptions on the Privacy Mechanism}

\bluechange{
Our model relies on the assumption that, upon receiving the lab's observation $Y$, Alice can remove the noise contribution $Z_\alpha = \sum_{i=1}^K \alpha_i Z_i$ and divide the result by $\alpha_0 > 0$ to obtain $X$.
This can be done because we assume that Alice knows the mixing coefficients $\alpha_0$, $\alpha_1$, \dots, $\alpha_K$ (she is the one doing the mixing) and that she knows the genotypes $Z_1,\dots,Z_K$ of the auxiliary noise individuals (which, as mentioned in Section~\ref{sec:intro}, already know their genotypes).
}

\bluechange{
However, the assumption that the noise individuals are willing to share their genotypes $Z_i$ with Alice can be a practical issue.
Alternatively, if the noise  individuals know their genotypes $Z_i$ but are not willing to share them with Alice, 
a secure multiparty computation scheme \cite{Yao1986, Goldreich1987, Lindell2020} 
can be used by the noise individuals to 
compute the noise $\sum_{i=1}^K \alpha_i Z_i$ only, and Alice can use the result to remove the noise contribution from $Y$.
If Alice does not want the noise individuals to learn $\sum_{i=1}^K \alpha_i Z_i$ in order to prevent them from colluding with the sequencing laboratory, Alice can create a random sequence of $t$ bits $(b_1, b_2, ..., b_t)$ where $t$ is the number of bits in the binary representation of $\sum_{i=1}^K \alpha_i Z_i$ that she wishes to compute.  She can then carry out a secure multiparty computation scheme with the noise individuals to compute $(c_1 \oplus b_1, c_2 \oplus b_2, ..., c_t \oplus b_t)$ where $(c_1, c_2, ..., c_t)$ is the sequence of bits in the binary representation of $\sum_{i=1}^K \alpha_i Z_i$.  She can then recover $(c_1, c_2, ..., c_t)$ using her knowledge of $(b_1, b_2, ..., b_t)$.  Since $(b_1, b_2, ..., b_t)$ is known only by Alice,  $\sum_{i=1}^K \alpha_i Z_i$ remains private from the noise individuals.
}


\bluechange{
In the case where the noise individuals do not know their genotypes, an alternative could be for 
Alice to prepare a second mixture only containing the noise $\sum_{i=1}^K \alpha_i Z_i$ and submit that for sequencing at a different lab.
Assuming that the labs are not colluding, this would allow Alice to subtract the noise term from $Y$ without knowledge of the $Z_i$s.
}

\bluechange{
Any of theses settings would lead to Alice being able to recover $X$, while the lab would only have access to $Y$. Hence, the goal of minimizing $I(X;Y)$ would be the same in all cases.
}

\subsection{Maximizing Privacy with a Utility Constraint}
\label{sec:utility_constraint}



\bluechange{
In practice, we would not be able mix the samples exactly according to proportions $\alpha_0, ..., \alpha_K$, and the sequencer does not always output the correct allele in every reading.  
Both of these issues manifest as noise in the observation $Y$, which can make reconstruction of $X$ by Alice more difficult when she subtracts $\sum_{k=1}^K \alpha_k Z_k$ from $Y$ to recover $X$. While the main information theoretic question we ask in (\ref{eq:main}) does not incorporate this noise, we can ensure that the proportion of Alice's sample in the mixture is large enough so that the reconstruction is robust against whatever noise is present.  This amounts to adding the constraint $\alpha_0 \geq \kappa$ 
to (\ref{eq:main}) for some constant $\kappa$ that ensures a level of utility of the scheme for reconstructing the Alice's allele.   This yields the following optimization problem:
\al{
\inf_{\alpha \in \R^{K+1} : \; \alpha > 0, \; \sum_{i=0}^K \alpha_i = 1, \; \alpha_0 \geq \kappa} 
I\left( X; \alpha_0X+Z_\alpha\right).
\label{eq:main_constrained}
}
Just as we obtained (\ref{eq:main2}), we can remove the constraint $\sum_{i=0}^K \alpha_i = 1$ from (\ref{eq:main_constrained}) to obtain
\al{
\inf_{\alpha \in \R^{K+1} : \; \alpha > 0, \; \alpha_0 /(\sum_{i=0}^K \alpha_i) \geq \kappa} 
I\left( X; \alpha_0X+Z_\alpha\right).
\label{eq:main_constrained2}
}
This can be thought of as restricting the signal to signal plus noise ratio of the mixture to be above a certain threshold.  Since this is an additional constraint on the optimization problem in (\ref{eq:main}), understanding (\ref{eq:main}) tells us the best possible privacy we can achieve for any choice of $\kappa$.  






We expect the ideas and techniques developed in this paper to provide useful insights for the additionally constrained problem in \eqref{eq:main_constrained2}.
We leave a careful analysis of this problem as future work, but here we provide some preliminary discussion. 
Since we have added a constraint, the lower bound we derive on (\ref{eq:main}) in Theorem \ref{thm:main_lower_bound} serves as a lower bound on (\ref{eq:main_constrained}) for any choice of $\kappa$.  
One can also try to modify the lower bound derived in Theorem~\ref{thm:main_lower_bound} to incorporate this additional constraint, which could lead to a larger lower bound. 
Notice that the constraint $\alpha_0 /(\sum_{i=0}^K \alpha_i) \geq \kappa$ is equivalent to the constraint $\sum_{i=1}^K \alpha_i \leq \alpha_0 (1 + \frac{1}{\kappa})$ which simply constrains the support of the noise distribution $\sum_{i=1}^K \alpha_i Z_i$ to have maximum support value $\alpha_0 (1 + \frac{1}{\kappa})$.  
By following the proof of Theorem \ref{thm:main_lower_bound}, we observe that   (\ref{eq:relax_const_bound}) with  $n = \alpha_0 (1 + \frac{1}{\kappa})$ serves as a lower bound on (\ref{eq:main_constrained2}).  However, when we plotted this support-constrained lower bound, we did not notice much of a difference compared to the original lower bound provided by Theorem \ref{thm:main_lower_bound}.


On the other hand, as expected, we did notice a difference when we compared the privacy of the optimal scheme in the unconstrained case (\ref{eq:main}) to the privacy of the constrained case (\ref{eq:main_constrained2}) for various parameter settings.  
It was observed however, that for many relevant parameter settings, the lower bound in Theorem \ref{thm:main_lower_bound} was still tight for a large range of $p$ relative to the optimal constrained solution.  An example of this is shown in Figure \ref{fig:constrained_opt}, where we see that the lower bound in Theorem \ref{thm:main_lower_bound} is tight compared to the constrained bound for $p < 0.3$.  

Another interesting observation is that if the constraint $\alpha_0/(\sum_{i=0}^K \alpha_i) \geq \kappa$ is added, it is often the case that using all $K$ noise individuals is not optimal.  With $K \leq 8$,  $\kappa = 0.05$,  and $\alpha_0 = 1$ as plotted in Figure \ref{fig:constrained_opt}, the optimal scheme for $p = 0.25$ is $\alpha = [1, 1, 1, 2, 2, 3, 4, 5, 0]$ i.e. the eighth noise individual is not used.  This is in contrast to the unconstrained case (\ref{eq:main_constrained}) where using more noise individuals always allows for increased privacy as discussed in Section \ref{sec:preliminaries}.
}


\begin{figure}[H]
\centering
\begin{tikzpicture}[scale=0.8]
\begin{axis}
[width=0.47\textwidth,
height=0.36\textwidth,
title ={Comparison of Bounds: $K \leq 8$},
ylabel = mutual information, 
xlabel = minor allele frequency (p),
ymax = .1,
legend pos = north east]

\addplot [
line width=1pt,
color=red!80!black
] table[x index=0,y index=1] {data/plt7_opt_constrained.dat};
\addlegendentry{optimal $\kappa = 0.05$}

\addplot [
line width=1pt,
color=green!80!black
] table[x index=0,y index=1] {data/plt7_bound.dat};
\addlegendentry{lower bound}

\end{axis}    
\end{tikzpicture}
\caption{
Mutual information for the optimal choice of $\alpha$ for $K \leq 8$ and $\kappa = 0.05$, 
and our lower bound from Theorem \ref{thm:main_lower_bound}.
The optimal constrained curve is computed using brute force search.}
\label{fig:constrained_opt}
\end{figure}


\subsection{Concluding Remarks}

Ensuring privacy for genetic data is an increasingly important topic, and the results in this paper can be seen as studying the fundamental limits of one particular privacy-preserving mechanism.
While there is room for extending our results to address various additional factors in the DNA sequencing pipeline, we characterize the best possible privacy that can be achieved by  mixing individuals in different proportions.
We do that by casting the problem as an information-theoretic worst-case noise characterization problem. 
Finally, we point out that this problem may be of independent interest as it adds to the literature on worst-case noise distributions, and it may have connections to other settings where information privacy is achieved through the addition of noise.





{\footnotesize

\bibliographystyle{ieeetr}
\bibliography{refs.bib}

}


\appendix
\label{sec:add_proofs}

\subsection{Simple Lower Bound on Mutual Information}
\label{sec:notzero}

When first considering the optimization problem in \eqref{eq:main}, one may intuitively expect the optimal value to be zero, since we can let $\alpha_0 \to 0$.
However, this is not the case due to the assumption that $\alpha > 0$ and the discrete nature of the random variables involved.
A simple way to see this is to let $Y = \alpha_0 X + \sum_{i=1}^K \alpha_i Z_i$ and notice that, if $Y=0$, it must be the case that $X=0$, since otherwise we would have $Y \geq \alpha_0$.
This observation allows us to lower bound $I(X;Y)$ as
\begin{align} 
    & I(X; Y) = \mathbb{E}_Y [D(p_{X|Y} \vert \vert \;  p_X)]
    \\ & = \sum_{y} p_Y(y) \sum_{x} p_{X|Y=y}(x) \log \left(\frac{p_{X|Y=y}(x)}{p_{X}(x)}\right)
    \\ & \geq p_Y(0) \left(p_{X|Y=0}(0) \log \left(\frac{p_{X|Y=0}(0)}{p_{X}(0)}\right) + p_{X|Y=0}(1) \log \left(\frac{p_{X|Y=0}(1)}{p_{X}(1)}\right)\right)
    \\ & = p_Y(0) p_{X|Y=0}(0) \log \left(\frac{p_{X|Y=0}(0)}{p_{X}(0)} \right) \label{eq:pxyzero}
    \\ & = (1-p)^K \cdot 1 \cdot \log \left(\frac{1}{1-p} \right)
    \\ & = (1-p)^K \log \left(\frac{1}{1-p} \right),
\end{align}
where in \eqref{eq:pxyzero} we used the fact that $p_{X|Y=0}(1) = 0$.
Observe that this lower bound is strictly greater than $0$ for $p > 0$ and does not depend on $\alpha_0$.  
Therefore, $I(X;Y)$ does not tend to zero as $\alpha_0 \to 0$.
As it turns out, the lower bound above is fairly loose, and Theorem~2 in our paper provides a much tighter lower bound.

\subsection{Auxiliary Lemmas} \label{sec:lemmas}

\begin{lemma}
    \label{lem:alpha0_condition}
    Fixing $\alpha_0 = 1$ and constraining the support of $Q$ to be integral in the optimization problem (\ref{eq:relax_inf1}) does not change the optimal value.
\end{lemma}
	
\begin{IEEEproof}
    Let $\alpha_0 \in \mathbb{N}.$ 
	Let $Q$ be any discrete random variable such that its pmf has minimum support value at $t = 0$.  
    Observe that
    \begin{align}
        & I(X ; \; \alpha_0 X +	Q) = I(X ; \; \frac{\alpha_0 X +Q}{\alpha_0}) = I(X ; \; X + D) \nonumber 
    \end{align} where $D = \frac{1}{\alpha_0}Q.$
    For a real number $x \in \R,$ we define $x \mod 1$ as $x - \floor{x}$ where $\floor{\cdot}$ is the floor function.
    Let $\hat{D} = D + S$,
	where $S = (-D) \mod{1}.$ 
	Then
	\begin{align} 
	    I(X ; \; \alpha_0 X + Q) & = I(X ; \;  X +	D) 
		\nonumber \\ & =  I(X ; \;  X + \hat{D} -  S)
		\nonumber \\ & \eqnum I(X ; \;  X + \hat{D}, \; S)
		\nonumber \\ & \geq I(X \; ; \;  X + \hat{D})
	\end{align} 
	where $(i)$ follows since from $X + \hat{D} - S,$ we can compute
	\begin{align}
	    & S = (-(X + \hat{D} - S)) \mod  1, \nonumber \\
	    &  X + \hat{D} = ( X + \hat{D} - S) + S. \nonumber
	\end{align}  
	Therefore it suffices to fix $\alpha_0=1$ and only consider discrete random variables $Q$ with integer support and minimum support value at $0$ in the optimization. 
\end{IEEEproof}

\begin{lemma}
	\label{lem:solve_KKT}
	A solution to the KKT conditions for (\ref{eq:relax_const_pert_posq}) is 
    \begin{align}
    \label{eq:KKT_conditions_sol_full}
    & q_{(i)} =  \beta (1 - \beta)^{i}  \text{ for } i \in \{0,...,n+1\} 
    \nonumber \\ & v_1 = - p \log \left( \frac{p}{(1-p)(1 - \beta) + p} \right) 
    \nonumber \\ & \quad - (1-p)\log \left( \frac{(1-p)(1 - \beta)}{(1-p)(1 - \beta) + p} \right)
    \nonumber \\ & v_2 = (1-p)\log \left( \frac{(1-p)(1 - \beta)}{(1-p)(1 - \beta) + p} \right) 
    \nonumber \\ & v_3 = p \log \left( \frac{p}{(1-p)(1 - \beta) + p} \right).
    \end{align}
	\end{lemma}
	\begin{IEEEproof}
	Let $f_0(q)$ be the objective function of (\ref{eq:relax_const_pert_posq}).  
The Lagrangian of (\ref{eq:relax_const_pert_posq}) is given in (\ref{eq:lagrangian}). 
The derivative of $f_0(q)$ with respect to $q_{(j)}$ is given by 
\begin{align}
& p \log \left(\frac{p q_{(j)}}{(1-p)q_{(j+1)} + p q_{(j)}} \right) \nonumber \\ & \quad \quad  + (1-p) \log \left( \frac{(1-p) q_{(j)}}{p q_{(j-1)} + (1-p) q_{(j)}} \right)
\end{align}  
for $j \in \{1, ...,  n\}$.
For $j=0$ and $j=n+1$, the derivative can be similarly computed
and the KKT conditions are 
\begin{align} 
& q_{(0)} = (1-p)^K,
\nonumber \\ & q_{(n+1)} = (1-p)^K (1 - (1-p)^K)^{n+1}, 
\nonumber \\ &     \sum_{j=0}^{n+1} q_{(j)} = 1 - (1 - (1-p)^K)^{n+2},
\nonumber \\ & \text{for $j = 1,...,n$,} \quad p \log \left(\frac{p q_{(j)}}{(1-p)q_{(j+1)} + p q_{(j)}} \right) 
\nonumber \\ & \quad + (1-p) \log \left( \frac{(1-p) q_{(j)}}{p q_{(j-1)} + (1-p) q_{(j)}} \right)+ v_1 = 0,
\nonumber \\ & p \log \left(\frac{p q_{(0)}}{(1-p)q_{(1)} + p q_{(0)}} \right) + v_1 + v_2 = 0,
\nonumber \\ & (1-p) \log \left( \frac{(1-p) q_{(n+1)}}{p q_{(n)} + (1-p) q_{(n+1)}} \right) + v_1 + v_3 = 0.
\label{eq:KKT}
\end{align}
The last three conditions can be rewritten as 
\begin{align} 
& \text{for $j = 1,...,n$,} \quad p \log \left(\frac{p }{(1-p)\frac{q_{(j+1)}}{q_{(j)}} + p } \right) \nonumber \\ & \quad \quad + (1-p) \log \left( \frac{(1-p) }{p \frac{q_{(j-1)}}{q_{(j)}} + (1-p) } \right)   + v_1  = 0, \nonumber \\ 
& p \log \left(\frac{p }{(1-p) \frac{q_{(1)}}{q_{(0)}} + p } \right)  + v_1 + v_2  = 0, \nonumber \\
& (1-p) \log \left( \frac{(1-p) }{p \frac{q_{(n)}}{q_{(n+1)}} + (1-p) } \right) + v_1 + v_3 = 0. 
\end{align}
Notice that, if the ratio $\frac{q_{(j+1)}}{q_{(j)}}$ between consecutive values of $q_{(j)}$ is the same for all $j$,  then $v_1$, $v_2$, $v_3$ can  be chosen so that these derivatives equal $0$ for all $j$.  
Setting \[\frac{q_{(j+1)}}{q_{(j)}} = (1 - (1-p)^K),\] 
we have that (\ref{eq:KKT_conditions_sol_full}) is
a solution to all equations in (\ref{eq:KKT}).
\end{IEEEproof}

\begin{lemma} \label{lem:lower_bound}
	For any $K\in \mathbb{N},$ $p \in [0, 0.5],$ and $\alpha_i > 0, \; i \in 0,1,...,K,$  we have that 
	\begin{align*} 
	I(X \; ; \; \alpha_0 X + \sum_{i=1}^{K} \alpha_i Z_i) \geq H(p) - 1 + p^{K+1} + (1-p)^{K+1}.
	\end{align*}
\end{lemma}


\begin{IEEEproof}
	Define $S_\alpha = \sum_{i=0}^{K} \alpha_i,$ and let $T_Y$ be the set of support values of $Y = \alpha_0 X + \sum_{i=1}^{K} \alpha_i Z_i.$  Clearly, $S_\alpha$ is the maximum element in $T_Y.$    We have that 
	\begin{align*} 
	& I(X \; ; \; Y) \\
	& = H(p) - H(X \; | \; Y) \\
	& \geq H(p) - (1 - p^{K+1} - (1-p)^{K+1}) \\
	& =  H(p) - 1 + p^{K+1} + (1-p)^{K+1}
	\end{align*}
	where the third line follows because
	\begin{align*} 
	& H(X \; | \; Y) \\
	& = \sum_{t \in T_Y} \Pr(Y = t) H(X \; | \; Y = t)  \\
	& = \Pr(Y = 0) \cdot H(X \; | \; Y = 0) 
	\\ & + \Pr(Y = S_\alpha) \cdot H(X \; | \; Y = S_\alpha) \\ 
	& + \sum_{t \in T_Y \setminus \{0, S_\alpha\}} \Pr(Y = t) H(X \; | \; Y = t) \\
	& = (1-p)^{K+1} \cdot 0 + p^{K+1} \cdot 0 \\ 
	& + \sum_{t \in T_Y \setminus \{0, S_\alpha\}} \Pr(Y = t) H(X \; | \; Y = t) \\
	& \leq \sum_{t \in T_Y \setminus \{0, S_\alpha\}} \Pr(X + \sum_{i=1}^{K} \alpha_i Z_i = t) \\
	& = 1 - p^{K+1} - (1-p)^{K+1}.
	\end{align*}
\end{IEEEproof}

\begin{lemma} \label{lem:binary_scheme_performance}
    For any $K \in \mathbb{N}, p \in [0, 0.5],$ we have that 
	\begin{align*} 
	& I(X \; ; \; X + \sum_{i=1}^{K} 2^{i-1} Z_i) \\ 
	& =  H(p) - \sum_{i=1}^K (p^i (1-p) + p (1-p)^i) H \left(\frac{1}{1 + \frac{p (1-p)^i}{p^i (1-p)}}\right).
	\end{align*}
\end{lemma}


\begin{IEEEproof}
We will prove this by induction on $K \in \mathbb{N}.$  Let $p \in [0, 0.5].$  For $K = 1,$ we have that 
\begin{align*} 
& I(X \; ; \; X + Z_1) \\ 
& = H(p) - H(X \; | \; X + Z_1) \\
& = H(p) - 2p(1-p)H(0.5)
\end{align*}		
which matches the formula.
Assume the formula holds for the $(K-1)$th  case where $K > 1.$  Consider the $K$th case: 
\begin{align*} 
& I\left(X \; ; \; X + \sum_{i=1}^{K} 2^{i-1} Z_i\right) \\
& = H(p) - H\left(X \; | \; X + \sum_{i=1}^{K} 2^{i-1} Z_i\right) \\
& = H(p) - (1-p)H\left(X \; | \; X + \sum_{i=1}^{K-1} 2^{i-1} Z_i\right) \\
& - p H\left(X \; | \; X + \sum_{i=1}^{K-1} 2^{i-1} Z_i\right) \\
& - (p^{K}(1-p) + p(1-p)^{K}) H\left(\frac{p^{K} (1-p)}{p^{K} (1-p) + p (1-p)^{K}}\right) \\
& = H(p) - H\left(X \; | \; X + \sum_{i=1}^{K-1} 2^{i-1} Z_i\right) \\ 
& - (p^{K}(1-p) + p(1-p)^{K}) H\left(\frac{1}{1 + \frac{p(1-p)^K}{p^K(1-p)}}\right) \\
& = H(p) - \sum_{i=1}^K (p^i(1-p) + p(1-p)^i) H\left(\frac{1}{1 + \frac{p(1-p)^i}{p^i(1-p)}}\right).
\end{align*}
\end{IEEEproof}

\begin{lemma} \label{lem:optimality_of_binary}
	For any $K \in \mathbb{N},$ the binary scheme is optimal for $p = 0.5.$
\end{lemma}


\begin{IEEEproof}
For any $K \in \mathbb{N}$ and $p = 0.5,$ the performance of the binary scheme is given by
\begin{align*}
	& I\left(X \; ; \; X + \sum_{i=1}^{K} 2^{i-1} Z_i\right) \\ 
	& =  H(0.5) - 2\sum_{i=1}^K (0.5)^{i+1} H(0.5) \\
	& = 1 - \sum_{i=1}^K (0.5)^{i} \\
	& = 1 - \left(\frac{1 - (0.5)^{K+1}}{0.5} - 1\right) \\
	& = (0.5)^{K} \\
	& = 1 - 1 + (0.5)^{K+1} + (0.5)^{K+1}
\end{align*}
Thus, the performance of the binary scheme matches the lower bound for $p = 0.5.$
\end{IEEEproof}

\begin{lemma} 
	\label{lem:uniform_schem_formula}
	For any $K \in \mathbb{N}, p \in [0, 0.5],$ we have that 
	\begin{align*} 
	& I\left(X \; ; \; X + \sum_{i=1}^K Z_i\right)
	\\ &  = H(p) - \sum_{i = 1}^{K} (1-p)^{K+1-i} p^{i} \binom{K+1}{i} H\left(\frac{i}{K+1}\right)
	\end{align*}
\end{lemma}

\begin{IEEEproof}
    We have that 
    \begin{align*} 
	& I\left(X \; ; \; X + \sum_{i=1}^K Z_i\right) 
	\\ & = H(p) - H\left(X \; | \; X + \sum_{i=1}^K Z_i\right) 
	\\ & = H(p) 
	\\ & - \sum_{t = 0}^{K+1} \Pr\left(X + \sum_{i=1}^K Z_i = t\right) H\left(X \; | \; X + \sum_{i=1}^K Z_i = t\right)
	\end{align*}
    and
    \[\Pr\left(X + \sum_{i=1}^K Z_i = t\right) = (1-p)^{K+1-t} p^{t} \binom{K+1}{t}.\] Due to Bayes rule, we have 
	\begin{align*}
	    \Pr\left(X = 1 \; | \; X + \sum_{i=1}^K Z_i = t\right) = \frac{t}{K+1},
	\end{align*}
    and thus, 
	\begin{align}
	    & H\left(X \; | \; X + \sum_{i=1}^K Z_i = t\right)
	    = H\left(\frac{t}{K+1}\right).
	\end{align}
    Thus,
    \begin{align*} 
	& I\left(X \; ; \; X + \sum_{i=1}^K Z_i\right) 
	\\ & = H(p) 
	\\ & - \sum_{t = 0}^{K+1} \Pr\left(X + \sum_{i=1}^K Z_i = t\right) H\left(X \; | \; X + \sum_{i=1}^K Z_i = t\right)
	\\ & = H(p) - \sum_{t = 0}^{K+1} (1-p)^{K+1-t} p^{t} \binom{K+1}{t} H\left(\frac{t}{K+1}\right)
    \\ & = H(p) - \sum_{t = 1}^{K} (1-p)^{K+1-t} p^{t} \binom{K+1}{t} H\left(\frac{t}{K+1}\right).
	\end{align*}
\end{IEEEproof}

\begin{lemma}
\label{lem:optimality_of_uniform}
For any $K \in \mathbb{N},$ there exists some $p^* > 0$ such that the uniform scheme is optimal for $p < p^*.$
\end{lemma}

\begin{IEEEproof}
    Consider the optimization problem (\ref{eq:main}) with $\alpha_0$ fixed to $1.$  We can assume this without loss of generality since scaling all $\alpha_i$s by the same constant does not change the mutual information.
	We have that 
	\begin{align*} 
	I(X \; ; \; X + \sum_{i=1}^K \alpha_i Z_i ) = H(p) - H(X \; | \; X + \sum_{i=1}^K \alpha_i Z_i)
	\end{align*} 
	and
	\begin{align} \label{eq:cond_ent_sum} 
	&H(X \; | \; X + \sum_{i=1}^K \alpha_i Z_i) \nonumber \\
	& = \sum_{t \in \N_0} \Pr(X  + Z_\alpha = t) H(X \; | \; X + Z_\alpha = t)
	\end{align} 
	Observe that as $p \to 0,$ we have that $H(p) \to 0,$ and $H(X \; | \; X + \sum_{i=1}^K \alpha_i Z_i) \to 0$ since $0 \leq H(X \; | \; X + \sum_{i=1}^K \alpha_i Z_i) \leq H(p).$  Thus, $I(X \; ; \; X + \sum_{i=1}^K \alpha_i Z_i ) \to 0$ as $p \to 0.$  We will show that each nonzero term in  (\ref{eq:cond_ent_sum}) decreases like $o(p),$ except the $t=1$ term which decreases like $\Theta(p)$ if there exists some $\alpha_i = 1$ for $i \geq 1.$  Therefore, we will show that the uniform scheme  maximizes the coefficient in the asymptotic expression for the $t=1$ term, proving that the uniform scheme is optimal as $p \to 0.$  
	
	For any $t$ such that $0 \leq t < 1,$ we have that $H(X \; | \; X + Z_\alpha = t) = 0$ because if $X = 1,$ then $X+Z_\alpha \geq 1.$
	
	Next, consider the term $\Pr(X + Z_\alpha = 1) H(X \; | \; X + Z_\alpha = 1).$  Observe that 
	\begin{align*} & \Pr(X + Z_\alpha = 1) \\& = (1-p) \Pr(Z_\alpha = 1) \\ & + p \Pr(Z_\alpha = 0) 
	\end{align*}
	and
	\begin{align*} 
	&\Pr(Z_\alpha = 1) = p (1-p)^{K-1}|\{i : \alpha_i = 1\}| + O(p^2), \\
	& \Pr(Z_\alpha = 0) = (1-p)^K,
	\end{align*}
	where the $O(p^2)$ term in expression for $\Pr(Z_\alpha = 1)$ follows because it is possible for 2 or more $Z_i$s to equal 1 such that the corresponding $\alpha_i$s add up to $1$.	If $|\{i : \alpha_i = 1\}| \geq 1,$ we have that 
	\begin{align} 
	&\Pr(X+Z_\alpha = 1) H(X \; | \; X+Z_\alpha = 1) \nonumber \\
	& = (p (1-p)^{K}(1 + |\{i : \alpha_i = 1\}|) + o(p)) 
	\nonumber \\& \times H(X \; | \; X+Z_\alpha = 1) \nonumber \\
	& = (p (1-p)^{K}(|\{i : \alpha_i = 1\}|) + o(p)) \nonumber \\
	& \times \log\left(\frac{p (1-p)^{K}(1+|\{i : \alpha_i = 1\}|)+ o(p)}{p (1-p)^{K}|\{i : \alpha_i = 1\}| + o(p)} \right) \nonumber \\ 
	& + p (1-p)^{K} \log\left(\frac{p (1-p)^{K}(1+|\{i : \alpha_i = 1\}|)+ o(p)}{p (1-p)^{K}}\right) \nonumber \\
	& = (p (1-p)^{K}(|\{i : \alpha_i = 1\}|) + o(p)) \nonumber \\
	& \times \log\left(\frac{(1+|\{i : \alpha_i = 1\}|)+ o(1)}{|\{i : \alpha_i = 1\}| + o(1)} \right) \nonumber \\ 
	& + p (1-p)^{K} \log((1+|\{i : \alpha_i = 1\}|)+ o(1)) \nonumber \\
	& \sim p \bigg(|\{i : \alpha_i = 1\}|\log \left(1 + \frac{1}{|\{i : \alpha_i = 1\}|}\right) \nonumber \\
	& +  \log(1+|\{i : \alpha_i = 1\}|)\bigg)
	\end{align}
	The coefficient of $p$ in the asymptotic expression for $\Pr(T(K) = 1) H(X \; | \; T(K) = 1)$ above is maximized when $|\{i : \alpha_i = 1\}| = K.$  This is because  
	\begin{align*} 
	\frac{d}{d x} \left(x \log \left(1 + \frac{1}{x}\right) + \log(1 + x)\right) > 0
	\end{align*} 
	for all $x > 0.$
		In contrast, suppose that $|\{i : \alpha_i = 1\}| = 0.$  Then, we have that 
    \begin{align} 
	&\Pr(Z_\alpha = 1) \nonumber \\
	& = b (1-p)^{K-a} p^a + o(p^a)	\end{align}
	for some $a \in \N$ such that $a \geq 2,$ and therefore, 
	\begin{align} 
	&\Pr(X+Z_\alpha = 1) H(X \; | \; X+Z_\alpha = 1) \nonumber \\
	& = (p(1-p)^K +b (1-p)^{K+1-a} p^a + o(p^a))  \nonumber \\ & \times H(X \; | \; X+Z_\alpha = 1) \nonumber \\
	& = (b (1-p)^{K+1-a} p^a + o(p^a)) \nonumber \\ & \times \log\left(\frac{(p(1-p)^K +b (1-p)^{K+1-a} p^a + o(p^a))}{b (1-p)^{K+1-a} p^a + o(p^a))} \right) \nonumber \\ 
	& + p (1-p)^{K} \log\left(\frac{(p(1-p)^K +b (1-p)^{K+1-a} p^a + o(p^a))}{p (1-p)^{K}}\right) \nonumber \\
	& = (b (1-p)^{K+1-a} p^a + o(p^a))  \log\left(1 + \frac{(1-p)^{a-1}}{b p^{a-1} + o(p^{a-1}))} \right) \nonumber \\
	& + p (1-p)^{K} \log\left((1 +b (1-p)^{1-a} p^{a-1} + o(p^{a-1}))\right) \nonumber \\
	& \sim b p^a   \log\left( \frac{1}{b p^{a-1}} \right) + p (1-p)^{K} \frac{b p^{a-1} }{\ln(2)} \nonumber \\
	& = o(p).
	\end{align}

	Now consider any $t>1.$  For such a value of $t,$ we have that \begin{align*} \Pr(Z_\alpha = t-1) = d (1-p)^{K-c} p^c  + o(p^c)
	\end{align*}  
	for some $c \in \mathbb{N}, \; c \geq 1$ and $d \in \mathbb{N}$ because there must be at least one $i \in [K]$ such that $Z_i=1$ in this case.  Furthermore, we have 
	\begin{align*} 
	\Pr(Z_\alpha = t) = b (1-p)^{K-a} p^a + o(p^a)
	\end{align*}  
	for some $a \in \mathbb{N}, \; a \geq 1$ and $b \in \mathbb{N}$ because there must be at least one $i \in [K]$ such that $Z_i=1$ in this case.  Thus, 
	\begin{align}
	\label{eq:entropy_term_t_bigger_1}
	&\Pr(X + Z_\alpha = t) H(X \; | \; X + Z_\alpha = t) \nonumber \\
	&= (d (1-p)^{K-c+1} p^{c} + o(p^{c})) \nonumber \\
	& \times \log(1 + \frac{ b (1-p)^{K-a} p^{a+1} ) + o(p^{a+1})}{d (1-p)^{K-c+1} p^{c} + o(p^{c})}) \nonumber \\
	& + (b (1-p)^{K-a} p^{a+1} + o(p^{a+1})) \nonumber \\
	& \times \log(1 + \frac{d (1-p)^{K-c+1} p^{c} + o(p^{c})}{b (1-p)^{K-a} p^{a+1} + o(p^{a+1})}) \nonumber \\
	& = o(p).
	\end{align}
    Thus, the only term in (\ref{eq:cond_ent_sum}) that decays like $\Theta(p)$ as $p \to 0$ is the $t=1$ term when there is at least one $i \in [K]$ such that $\alpha_i = 1$. Furthermore,  the uniform scheme maximizes the coefficient for this term.
    Thus, there exists some $p^* > 0$ such that the uniform scheme is optimal for $p < p^*.$
	
	We now justify the last line in (\ref{eq:entropy_term_t_bigger_1}). 
	Consider the first term above in the second line of (\ref{eq:entropy_term_t_bigger_1}).  If $c < a+1,$ it follows that 
	\begin{align*}
	&(d (1-p)^{K-c+1} p^{c} + o(p^{c})) \\
	& \times \log\left(1 + \frac{ b (1-p)^{K-a} p^{a+1} ) + o(p^{a+1})}{d (1-p)^{K-c+1} p^{c} + o(p^{c})}\right) \\
	& = (d (1-p)^{K-c+1} p^{c} + o(p^{c})) \\
	&\times \log\left(1 + \frac{ b (1-p)^{K-a} p^{a+1-c} + o(p^{a+1-c})}{d (1-p)^{K-c+1} + o(1)}\right) \\
	& \sim (d (1-p)^{K-c+1} p^{c} + o(p^{c})) \\ 
	& \times \frac{ b (1-p)^{K-a} p^{a+1-c} + o(p^{a+1-c})}{\ln(2)(d (1-p)^{K-c+1} + o(1))} \\
	& = o(p).
	\end{align*}
	If $c \geq a+1,$ it follows that 
	\begin{align*}
	& (d (1-p)^{K-c+1} p^{c} + o(p^{c})) \\
	& \times \log\left(1 + \frac{ b (1-p)^{K-a} p^{a+1}  + o(p^{a+1})}{d (1-p)^{K-c+1} p^{c} + o(p^{c})}\right)\\
	& \leq (d (1-p)^{K-c+1} p^{c} + o(p^{c})) \\
	& \times \log\left(\frac{1}{d (1-p)^{K-c+1} p^{c} + o(p^{c})}\right)\\
	& \leq (d (1-p)^{K-c+1} p^{c} + o(p^{c}))\log\left(\frac{1}{(1-p)^{K} p^{K}}\right)\\
	& = K (d (1-p)^{K-c+1} p^{c} + o(p^{c})) \\
	& \times \log\left(\frac{1}{p(1-p)}\right)\\
	& = o(p).
	\end{align*}
	Finally, consider the second term.  It follows that
	\begin{align*}	
	& (b (1-p)^{K-a} p^{a+1} + o(p^{a+1})) \\
	& \times \log\left(1 + \frac{d (1-p)^{K-c+1} p^{c} + o(p^{c})}{b (1-p)^{K-a} p^{a+1} + o(p^{a+1})}\right) \\
	& \leq  (b (1-p)^{K-a} p^{a+1} + o(p^{a+1})) \\
	& \times \log\left(\frac{1}{b (1-p)^{K-a} p^{a+1} + o(p^{a+1})}\right)  \\
	& \leq  (b (1-p)^{K-a} p^{a+1} + o(p^{a+1})) \\
	& \times \log\left(\frac{1}{ (1-p)^{K} p^{K} }\right)  \\
	& =  K(b (1-p)^{K-a} p^{a+1} + o(p^{a+1})) \\
	& \times \log\left(\frac{1}{ p (1-p) }\right) \\
	& = o(p).
	\end{align*}
\end{IEEEproof}
\begin{lemma}
\label{lem:main_recursion}
    For $M,N \in \mathbb{N}$ such that $M>0$ and $N>1,$ we have that 
	\begin{align*} 
		& H(\tilde{X} \; | \; T_{M,N})  
		\\ & = H(\tilde{X} \; | \; T_{M,N-1}) 
		\\ & - p \!\!\!\! \sum_{t=-S_{M,N-1}}^{-S_{M,N-1}+2A_M} \!\!\!\!\!\!\!\!\!\! \Pr(T_{M,N-1} = t) H(\tilde{X} \; | \; T_{M,N-1} = t)
		\\ & - 	(1-p) \!\!\!\! \sum_{t=S_{M,N-1}-2A_M}^{S_{M,N-1}} \!\!\!\!\!\!\!\!\!\! \Pr(T_{M,N-1} = t) H(\tilde{X} \; | \; T_{M,N-1} = t)
		\\ & + \sum_{t=-A_M}^{A_M} \Pr(T_{M,N} = t) H(\tilde{X} \; | \; T_{M,N} = t). 
	\end{align*}    
\end{lemma}

\begin{IEEEproof}
For $N \geq 2,$ we have that 
\begin{align*} 
& H(\tilde{X} \; | \; T_{M, N}) 
\\ & = 
\sum_{t=-S_{M,N}}^{S_{M,N}} \Pr(T_{M, N} = t) H(\tilde{X} \; | \; T_{M, N} = t) 
\\ & = 
\sum_{t=-S_{M,N}}^{-A_M-1} \Pr(T_{M, N} = t) H(\tilde{X} \; | \; T_{M, N} = t) 
\\ & + 
\sum_{t=A_M+1}^{S_{M,N}} \Pr(T_{M, N} = t) H(\tilde{X} \; | \; T_{M, N} = t) 
\\ & + 
\sum_{t=-A_M}^{A_M} \Pr(T_{M, N} = t) H(\tilde{X} \; | \; T_{M, N} = t) 
\\ & = \sum_{t=-S_{M,N-1}}^{S_{M,N-1}-2A_M-1} \! \! \!\!\!\!\!\!\!\! (1-p) \Pr(T_{M, N-1} = t) H(\tilde{X} \; | \; T_{M, N-1} = t)
\\ & + \sum_{t=-S_{M,N-1}+2A_M+1}^{S_{M,N-1}} \! \! \!\!\!\!\!\!\!\! p \Pr(T_{M, N-1} = t) H(\tilde{X} \; | \; T_{M, N-1} = t) 
\\ & + 
\sum_{t=-A_M}^{A_M} \Pr(T_{M, N} = t) H(\tilde{X} \; | \; T_{M, N} = t) 
\\ & = 
\sum_{t=-S_{M,N-1}+2A_M+1}^{S_{M,N-1}-2A_M-1} \Pr(T_{M, N-1} = t) H(\tilde{X} \; | \; T_{M, N-1} = t)
\\ & + 
\sum_{t=-S_{M,N-1}}^{-S_{M,N-1}+2A_M} \!\!\!\!  \!\!\!\!  (1-p) \Pr(T_{M, N-1} = t) H(\tilde{X} \; | \; T_{M, N-1} = t)
\\ & + 
\sum_{t=S_{M,N-1}-2A_M}^{S_{M,N-1}} p\Pr(T_{M, N-1} = t) H(\tilde{X} \; | \; T_{M, N-1} = t)
\\ & + 
\sum_{t=-A_M}^{A_M} \Pr(T_{M, N} = t) H(\tilde{X} \; | \; T_{M, N} = t) 
\\ & = 
H(\tilde{X} \; | \; T_{M, N-1}) 
\\ & - 
p \sum_{t=-S_{M,N-1}}^{-S_{M,N-1}+2A_M} \Pr(T_{M, N-1} = t) H(\tilde{X} \; | \; T_{M, N-1} = t)
\\ & - 
(1-p) \sum_{t=S_{M,N-1}-2A_M}^{S_{M,N-1}} \!\!\!\!\!\!\!\!  \Pr(T_{M, N-1} = t) H(\tilde{X} \; | \; T_{M, N-1} = t)
\\ & + 
\sum_{t=-A_M}^{A_M} \Pr(T_{M, N} = t) H(\tilde{X} \; | \; T_{M, N} = t)
\end{align*}
where the third equality follows from the fact that if $t < -A_M,$ 
\begin{align*} 
& H(\tilde{X} \; | \; T_{M, N} = t)
\\ & = 
H(\Pr(\tilde{X} = 1 \; | \; T_{M, N} = t))
\\ & = 
H\left(\frac{p\Pr(T_{M, N} = t \; | \; \tilde{X} = 1) }{\Pr(T_{M, N} = t)}\right)
\\ & = 
H\left(\frac{1}{1 + \frac{(1-p)\Pr(T_{M, N}-\tilde{X} = t +1)}{p\Pr(T_{M, N}-\tilde{X} = t -1)}}\right) 
\\ & = 
H\left(\frac{1}{1 + \frac{(1-p)^2\Pr(T_{M, N-1}-\tilde{X} = t +1+2^{N-1})}{(1-p)p\Pr(T_{M, N-1}-\tilde{X} = t -1+2^{N-1})}}\right) 
\\ & = 
H\left(\frac{1}{1 + \frac{(1-p)\Pr(T_{M, N-1}-\tilde{X} = t +1+2^{N-1})}{p\Pr(T_{M, N-1}-\tilde{X} = t -1+2^{N-1})}}\right) 
\\ & = 
H\left(\frac{p\Pr(T_{M, N-1} = t + 2^{N-1} \; | \; \tilde{X} = 1) }{\Pr(T_{M, N-1} = t + 2^{N-1})}\right)
\\ & =	 
H\left(\tilde{X} \; | \; T_{M, N-1} = t+ 2^{N-1}\right), 
\end{align*}
and if $t > A_M,$
\begin{align*}  
H(\tilde{X} \; | \; T_{M, N} = t) = 	H(\tilde{X} \; | \; T_{M, N-1} = t - 2^{N-1}).
\end{align*}
\end{IEEEproof}

\longversion{
\subsection{Explicit computation of (\ref{eq:closedform})}
\iscomment{add this in long version}
}

\newpage

\longversion{

\section{Efficiently Computable Schemes}

Theorem~\ref{thm:main_lower_bound} provides a lower bound to the 
mutual information between $X$ and the lab's observation $Y$ for any mixing coefficients $\alpha_0,...,\alpha_K$, thus providing a bound to the privacy levels that can be achieved.
However,

\rdcomment{this may sound stupid, but i think it warrants a sentence here. `Theorem 1 gives us a lower bound on the mutual information between $X$ and ..., which would be the mixing ratio that provides the highest level of privacy. However, it does not provide any mixing scheme for approaching this lower bound.'}

While much faster than brute force search, the greedy algorithm used to obtain the upper bound requires $\Omega(2^K)$ time in the worst case.  At a high level, this is due to the fact that at the $j$th  step in the algorithm, computation of $I(X \; ; \; X + \alpha Z_{j} + \sum_{i = 1}^{j-1} \alpha_i Z_i)$ requires use of each support value in the pmf of $\alpha Z_{j} + \sum_{i = 1}^{j-1} \alpha_i Z_i,$ and there are $\Omega(2^j)$ support values in the worst case.  This motivates the design of efficiently computable schemes, which can then be evaluated using the lower bound in Theorem \ref{thm:main_lower_bound}.

Perhaps the most natural approach is given by the uniform scheme, which we define to set $\alpha_i = 1$ for all $i \in [K]\cup\{0\}.$  This is the scheme used by the authors of \cite{Maddah-Ali} to obtain their results.  The scheme is explicitly defined given $K,$ so no algorithm is needed to compute it.  The performance of the scheme for a given $p$ and $K$ can be computed in $O(K)$ time using the formula in Lemma \ref{lem:uniform_schem_formula} which is given in the appendix.

The uniform scheme is important because it is optimal as $p \to 0$ as proved in Lemma \ref{lem:optimality_of_uniform}
and exemplified by the optimal curve in Figure \ref{fig:optimal_scheme_K5_minor_allele}.

\begin{lemma}
\label{lem:optimality_of_uniform}
For any $K \in \mathbb{N},$ there exists some $p^* > 0$ such that the uniform scheme is optimal for $p < p^*.$
\end{lemma}


At $p= 0.5,$ in Figure \ref{fig:optimal_scheme_K5_minor_allele}, the optimal solution is given by the binary scheme, which we define to set $\alpha_0 = 1$ and $\alpha_i = 2^{i-1}$ for all $i \in [K].$  Similar to the uniform scheme, the binary scheme is explicitly defined given $K.$  Also, similar to the uniform scheme, there exists a formula for computing its performance given $p$ and $K$ in $O(K)$ time.  This formula is given in Lemma \ref{lem:binary_scheme_performance} in the appendix.  It is interesting that such a formula exists because the pmf of $\sum_{i=1}^K 2^{i-1} Z_i$ has $2^K$ support values. 

Most importantly, the optimality of the binary scheme at $p = 0.5$ for $K=5$ shown in Figure \ref{fig:optimal_scheme_K5_minor_allele} generalizes to all $K$ as proved in Lemma \ref{lem:optimality_of_binary}.

\begin{lemma} \label{lem:optimality_of_binary}
	For any $K \in \mathbb{N},$ the binary scheme is optimal for $p = 0.5.$
\end{lemma}

Because the uniform scheme is optimal for $p \to 0$ and the binary scheme is optimal for $p = 0.5,$  it is natural to try and combine these schemes to interpolate the performance for $p$ in the range $0<p<0.5.$  Combining only the uniform scheme and the binary scheme does not produce a scheme that approaches the lower bound.  However, the general shape of the lower bound can be captured in polynomial time by combining the uniform scheme, the linear scheme, and the binary scheme where the linear scheme is defined as $\alpha_0 = 1$ and $\alpha_i = i$ for $i \in [K].$  We define the $(K,L,U)$-binary-linear-uniform scheme to set $\alpha_0 = 1$ and 
\begin{align} &\alpha_i = 1 \quad i \in \{1, ..., U\} \nonumber \\  &\alpha_{U+i} = i \quad i \in \{1, ..., L\} \nonumber \\
&\alpha_{U+L+i} = 2^{i-1} \quad i \in \{1, ..., K-U-L\}
\end{align}


For a given $K$ and $p,$ we define the $K$-binary-linear-uniform scheme as the $(K,L^*,U^*)$-binary-linear-uniform scheme where
\begin{align} & (L^*, U^*) = \argmin_{(L,U) \in ([K]\cup\{0\})^2 \; : \; L + U \leq K} \\ &  I\left(X \; ; \; X + \sum_{i=1}^{U} Z_i + \sum_{j=1}^{L} j Z_{U+j} +  \sum_{k=1}^{K-U-L} 2^{k -1} Z_{U+L+k}\right) \nonumber.
\end{align}
The $K$-binary-linear-uniform scheme is computed in $O(K^4)$ time using a generalization of the recursion used to prove Lemma \ref{lem:binary_scheme_performance}.  

We compare the performance of the uniform scheme, the binary scheme, the $K$-binary-linear-uniform scheme, and the scheme generated by the greedy algorithm in Figure \ref{fig:efficiently_computable_schemes}.  We have observed that the scheme generated by the greedy algorithm is much tighter than the   $K$-binary-linear-uniform scheme after the initial hump in the lower bound as $K$ increases.

\begin{figure}[h]
\label{fig:efficiently_computable_schemes}
\centering
\begin{tikzpicture}[scale=0.8]
\begin{axis}
[width=0.47\textwidth,
height=0.36\textwidth,
title ={Comparison of Bounds: $K = 15$},
ylabel = mutual information, 
xlabel = minor allele frequency (p),
ymax = .6,
legend pos = north east]

\addplot [
line width=1pt,
color=red!80!black
] table[x index=0,y index=1] {data/plt2_greedy.dat};
\addlegendentry{greedy}

\addplot [
line width=1pt,
color=green!80!black
] table[x index=0,y index=1] {data/plt3_uniform.dat};
\addlegendentry{uniform}

\addplot [
line width=1pt,
color=purple!80!black
] table[x index=0,y index=1] {data/plt3_linear.dat};
\addlegendentry{linear}

\addplot [
line width=1pt,
color=orange!80!black
] table[x index=0,y index=1] {data/plt3_binary.dat};
\addlegendentry{binary}

\addplot [
line width=1pt,
color=yellow!80!black,
dashed
] table[x index=0,y index=1] {data/plt3_binary_linear_uniform.dat};
\addlegendentry{binary-linear-uniform}

\addplot [
line width=1pt,
color=blue!80!black,
dotted
] table[x index=0,y index=1] {data/plt2_bound.dat};
\addlegendentry{lower bound}

\end{axis}    
\end{tikzpicture}

\begin{tikzpicture}[scale=0.8]
\begin{axis}
[width=0.47\textwidth,
height=0.36\textwidth,
title ={Comparison of Bounds: $K = 15$},
ylabel = mutual information, 
xlabel = minor allele frequency (p),
ymax = .01]

\addplot [
line width=1pt,
color=red!80!black
] table[x index=0,y index=1] {data/plt2_greedy.dat};

\addplot [
line width=1pt,
color=green!80!black
] table[x index=0,y index=1] {data/plt3_uniform.dat};

\addplot [
line width=1pt,
color=purple!80!black
] table[x index=0,y index=1] {data/plt3_linear.dat};

\addplot [
line width=1pt,
color=orange!80!black
] table[x index=0,y index=1] {data/plt3_binary.dat};

\addplot [
line width=1pt,
color=yellow!80!black,
dashed
] table[x index=0,y index=1] {data/plt3_binary_linear_uniform.dat};

\addplot [
line width=1pt,
color=blue!80!black,
dotted
] table[x index=0,y index=1] {data/plt2_bound.dat};

\end{axis}    
\end{tikzpicture}
\caption{Comparison between all upper bounds presented in this paper and the lower bound from \ref{} for $K = 15$.}
\end{figure}

\section{Appendix}

\begin{lemma} 
	For any $K \in \mathbb{N}, p \in [0, 0.5],$ we have that 
	\begin{align*} 
	& I(X \; ; \; X + \sum_{i=1}^K Z_i)
	\\ &  = H(p) - \sum_{i = 1}^{K} p^{K+1-i} (1-p)^{i} \binom{K+1}{i} H\left(\frac{i}{K+1}\right)
	\end{align*}
	\label{lem:uniform_schem_formula}
\end{lemma}

\begin{lemma} \label{lem:binary_scheme_performance}
    For any $K \in \mathbb{N}, p \in [0, 0.5],$ we have that 
	\begin{align*} 
	& I(X \; ; \; X + \sum_{i=1}^{K} 2^{i-1} Z_i) \\ 
	& =  H(p) - \sum_{i=1}^K (p^i (1-p) + p (1-p)^i) H \left(\frac{1}{1 + \frac{p (1-p)^i}{p^i (1-p)}}\right).
	\end{align*}
\end{lemma}

\begin{lemma} \label{lem:lower_bound}
	For any $K\in \mathbb{N},$ $p \in [0, 0.5],$ and $\alpha_i \in \mathbb{N}, \; i \in [K],$  we have that 
	\begin{align*} 
	I(X \; ; \; X + \sum_{i=1}^{K} \alpha_i Z_i) \geq H(p) - 1 + p^{K+1} + (1-p)^{K+1}.
	\end{align*}
\end{lemma}

\begin{theorem} \label{theorem:main_recursion}
	Let
    \begin{align*} 
    T_{M, N} = X + \sum_{i=1}^M Z_i +  \sum_{j=1}^{N} 2^{j-1} Z_{M+j}
    \end{align*}
    and
    \begin{align*} 
    S_{M,N} = 1 + M + \sum_{i=1}^{N} 2^{i-1}.
    \end{align*}
    For $M,N \in \mathbb{N}$ such that $M>0$ and $N>1,$ we have that 
	\begin{align*} 
		& H(X \; | \; T_{M,N})  
		\\ & = H(X \; | \; T_{M,N-1}) 
		\\ & - (1-p) \!\!\!\! \sum_{t=-S_{M,N-1}}^{-S_{M,N-1}+2M} \!\!\!\!\!\!\!\!\!\! \Pr(T_{M,N-1} = t) H(X \; | \; T_{M,N-1} = t)
		\\ & - 	p \!\!\!\! \sum_{t=S_{M,N-1}-2M}^{S_{M,N-1}} \!\!\!\!\!\!\!\!\!\! \Pr(T_{M,N-1} = t) H(X \; | \; T_{M,N-1} = t)
		\\ & + \sum_{t=-M}^{M} \Pr(T_{M,N} = t) H(X \; | \; T_{M,N} = t). 
	\end{align*}
\end{theorem}

\begin{algorithm} \label{alg:uniform_binary_polytime}
	\SetAlgoLined
	\KwData{$N,$ $M,$ $p$}
	\KwResult{$I_{M,N}$}
	$n_1 \leftarrow \min(N, \ceil{\log(2M+2)})$\;
	$D_{n_1} \leftarrow$ array containing PMF for $\sum_{i=1}^M Z_i +  \sum_{j=1}^{n_1} 2^{j-1} Z_{M+j}$ starting at the lowest support value and ending at highest support value\;
	$D_{\text{low}} \leftarrow$ array containing first $2M+1$ entries of  $D_{n_1}$\;
	$D_{\text{high}} \leftarrow$ array containing last $2M+1$ entries of $D_{n_1}$\;
	$H_{n_1}  \leftarrow $ $\sum_{t=-S_{M,n_1}}^{S_{M,n_1}} \Pr(T_{M, n_1} = t) H(X \; | \; T_{M, n_1} = t)$ computed using $D_{n_1}$\; 
	$H_{\text{low}} \leftarrow $ 
	$\sum_{t=-S_{M,n_1}}^{-S_{M,n_1}+2M} \Pr(T_{M, n_1} = t) H(X \; | \; T_{M, n_1} = t)$ computed using $D_{\text{low}}$\; 
	$H_{\text{high}} \leftarrow $ $\sum_{t=S_{M,n_1}-2K}^{S_{M,n_1}} \Pr(T_{M, n_1} = t) H(X \; | \; T_{M, n_1} = t)$ computed using $D_{\text{high}}$\; 	
	$n \leftarrow n_1 + 1$\;
	\While{$n \leq N$}{
		$H_{n}  \leftarrow H_{n-1}$\;
		$H_{n} \leftarrow H_{n} - (1-p) H_{\text{low}} - p H_{\text{high}}$\;
		$D_{\text{low ext}} \leftarrow $ $D_{\text{low}}$ appended with $[0, \; 0]$\;
		$D_{\text{high ext}} \leftarrow $  $D_{\text{high}}$ prepended with $[0,\;  0]$\;
		$H_{\text{center}} \leftarrow$ 	$\sum_{t=-M}^{M} \Pr(T_{M, n} = t) H(X \; | \; T_{M, n} = t)$ computed using $p D_{\text{high ext}} + (1-p) D_{\text{low ext}}$\;
		$H_{n} \leftarrow H_{n} + H_{\text{center}}$\;
		$H_{\text{low}} \leftarrow p H_{\text{low}}$\; 
		$H_{\text{high}} \leftarrow  (1-p) H_{\text{high}}$\; 
		$D_{\text{low}} \leftarrow p D_{\text{low}}$\;
		$D_{\text{high}} \leftarrow (1-p) D_{\text{high}}$\;
		$n \leftarrow n + 1$\;
	}
	$I_{M,N} \leftarrow H(p) - H_N$

	\caption{Computation of $I(X \; ; \; X + \sum_{i=1}^M Z_i +  \sum_{j=1}^{N} 2^{j-1} Z_{M+j})$}
\end{algorithm}

\subsection{Proof of Lemma \ref{lem:uniform_schem_formula}}
    We have that 
    \begin{align*} 
	& I\left(X \; ; \; X + \sum_{i=1}^K Z_i\right) 
	\\ & = H(p) - H\left(X \; | \; X + \sum_{i=1}^K Z_i\right) 
	\\ & = H(p) 
	\\ & - \sum_{t = 0}^{K+1} \Pr\left(X + \sum_{i=1}^K Z_i = t\right) H\left(X \; | \; X + \sum_{i=1}^K Z_i = t\right).
	\end{align*}
     Thus,
     \[\Pr\left(X + \sum_{i=1}^K Z_i = t\right) = p^{K+1-t} (1-p)^{t} \binom{K+1}{t}.\] Due to Bayes rule, we have 
	\begin{align*}
	    \Pr\left(X = 1 \; | \; X + \sum_{i=1}^K Z_i = t\right) = \frac{t}{K+1},
	\end{align*}
    and thus, 
	\begin{align}
	    & H\left(X \; | \; X + \sum_{i=1}^K Z_i = t\right)
	    = H\left(\frac{t}{K+1}\right).
	\end{align}
    Thus,
    \begin{align*} 
	& I(X \; ; \; X + \sum_{i=1}^K Z_i) 
	\\ & = H(p) 
	\\ & - \sum_{t = 0}^{K+1} \Pr\left(X + \sum_{i=1}^K Z_i = t\right) H\left(X \; | \; X + \sum_{i=1}^K Z_i = t\right)
	\\ & = H(p) - \sum_{t = 0}^{K+1} p^{K+1-t} (1-p)^{t} \binom{K+1}{t} H\left(\frac{t}{K+1}\right)
    \\ & = H(p) - \sum_{t = 1}^{K} p^{K+1-t} (1-p)^{t} \binom{K+1}{t} H\left(\frac{t}{K+1}\right)
	\end{align*}


\subsection{Proof of Lemma \ref{lem:optimality_of_uniform}}
	Let $K \in \mathbb{N}$ and
	\begin{align*}
	&S_K = 1 + \sum_{i=1}^K \alpha_i, \quad \quad T(K) = X + \sum_{i=1}^K \alpha_i Z_i.
	\end{align*}
	We have that 
	\begin{align*} 
	I(X \; ; \; X + \sum_{i=1}^K \alpha_i Z_i ) = H(p) - H(X \; | \; X + \sum_{i=1}^K \alpha_i Z_i)
	\end{align*} 
	and
	\begin{align*} 
	&H(X \; | \; X + \sum_{i=1}^K \alpha_i Z_i) \\
	& = \sum_{t=-S_K}^{S_K} \Pr(T(K) = t) H(X \; | \; T(K) = t)
	\end{align*} 
	Observe that as $p \to 0,$ the terms of $ H(X \; | \; X + \sum_{i=1}^K \alpha_i Z_i)$ with the lowest powers of $p$ dominate.
	The only term containing no power of $p$ is  
	\begin{align*} & \Pr(T(K) = S_K) H(X \; | \; T(K) = S_K) \\ & = (1-p)^{K+1} H(X \; | \; T(K) = S_K)  = 0. 
	\end{align*}
	Next, consider the term $\Pr(T(K) = S_K-2) H(X \; | \; T(K) = S_K-2).$  Observe that 
	\begin{align*} & \Pr(T(K) = S_K-2) \\& = (1-p) \Pr(T(K)-X = S_K-3) \\ & + p \Pr(T(K)-X = S_K-1) 
	\end{align*}
	and
	\begin{align*} 
	&\Pr(T(K)-X = S_K-3) = p (1-p)^{K-1}|\{i : \alpha_i = 1\}|, \\
	& \Pr(T(K)-X = S_K-1) = (1-p)^K. 
	\end{align*}
	We than have that 
	\begin{align*} 
	&\Pr(T(K) = S_K-2) H(X \; | \; T(K) = S_K-2) \\
	& = p (1-p)^{K}(1 + |\{i : \alpha_i = 1\}|) H(X \; | \; T(K) = S_K-2) \\
	& = p (1-p)^{K}(\{i : \alpha_i = 1\}| \\
	& \times \log(\frac{p (1-p)^{K}(1+|\{i : \alpha_i = 1\}|)}{p (1-p)^{K}|\{i : \alpha_i = 1\}|})) \\ 
	& + p (1-p)^{K} \log(\frac{p (1-p)^{K}(1+|\{i : \alpha_i = 1\}|)}{p (1-p)^{K}}) \\
	& = p (1-p)^{K}(\{i : \alpha_i = 1\}|\log(\frac{1+|\{i : \alpha_i = 1\}|}{|\{i : \alpha_i = 1\}|})) \\
	& + p (1-p)^{K} \log(1+|\{i : \alpha_i = 1\}|)
	\end{align*}
	The above function is maximized when $|\{i : \alpha_i = 1\}| = K.$  This is because  
	\begin{align*} 
	\frac{d}{d x} (x \log(1 + \frac{1}{x}) + \log(1 + x)) > 0
	\end{align*} 
	for all $x > 0.$
	
	Now consider any $t$ such that $t < S_K-2.$  For such a value of $t,$ we have that \begin{align*} \Pr(T(K) - X = t-1) = d (1-p)^{K-c} p^c  + o(p^c)
	\end{align*}  
	for some $c \in \mathbb{N}, \; c \geq 1$ and $d \in \mathbb{N}$ because there must be at least one $i \in [K]$ such that $Z_i=-1$ in this case.  Furthermore, we have 
	\begin{align*} 
	\Pr(T(K) - X = t+1) = b (1-p)^{K-a} p^a + o(p^a)
	\end{align*}  
	for some $a \in \mathbb{N}, \; a \geq 1$ and $b \in \mathbb{N}$ because there must be at least one $i \in [K]$ such that $Z_i=-1$ in this case.  Thus, 
	\begin{align*}     
	&\Pr(T(K) = t) H(X \; | \; T(K) = t) \\ 
	&= (d (1-p)^{K-c+1} p^{c} + o(p^{c})) \\
	& \times \log(1 + \frac{ b (1-p)^{K-a} p^{a+1} ) + o(p^{a+1})}{d (1-p)^{K-c+1} p^{c} + o(p^{c})}) \\
	& + (b (1-p)^{K-a} p^{a+1} + o(p^{a+1})) \\
	& \times \log(1 + \frac{d (1-p)^{K-c+1} p^{c} + o(p^{c})}{b (1-p)^{K-a} p^{a+1} + o(p^{a+1})}) \\
	& = o(p).
	\end{align*}

	Thus, there exists some $\epsilon > 0$ such that the uniform scheme is optimal for $p < \epsilon.$
	
	We now justify the last line in the equation above.   
	Consider the first term above in the second line.  If $c < a+1,$ it follows that 
	\begin{align*}
	&(d (1-p)^{K-c+1} p^{c} + o(p^{c})) \\
	& \times \log(1 + \frac{ b (1-p)^{K-a} p^{a+1} ) + o(p^{a+1})}{d (1-p)^{K-c+1} p^{c} + o(p^{c})}) \\
	& = (d (1-p)^{K-c+1} p^{c} + o(p^{c})) \\
	&\times \log(1 + \frac{ b (1-p)^{K-a} p^{a+1-c} + o(p^{a+1-c})}{d (1-p)^{K-c+1} + o(1)}) \\
	& = (d (1-p)^{K-c+1} p^{c} + o(p^{c})) \\ 
	& \times \frac{ b (1-p)^{K-a} p^{a+1-c} + o(p^{a+1-c})}{d (1-p)^{K-c+1} + o(1)} \\
	& = o(p).
	\end{align*}
	If $c \geq a+1,$ it follows that 
	\begin{align*}
	& (d (1-p)^{K-c+1} p^{c} + o(p^{c})) \\
	& \times \log(1 + \frac{ b (1-p)^{K-a} p^{a+1}  + o(p^{a+1})}{d (1-p)^{K-c+1} p^{c} + o(p^{c})})\\
	& \leq (d (1-p)^{K-c+1} p^{c} + o(p^{c})) \\
	& \times \log(\frac{1}{d (1-p)^{K-c+1} p^{c} + o(p^{c})})\\
	& \leq (d (1-p)^{K-c+1} p^{c} + o(p^{c}))\log(\frac{1}{(1-p)^{K} p^{K}})\\
	& = K (d (1-p)^{K-c+1} p^{c} + o(p^{c})) \\
	& \times \log(\frac{1}{p(1-p)})\\
	& = o(p).
	\end{align*}
	Finally, consider the second term.  It follows that
	\begin{align*}	
	& (b (1-p)^{K-a} p^{a+1} + o(p^{a+1})) \\
	& \times \log(1 + \frac{d (1-p)^{K-c+1} p^{c} + o(p^{c})}{b (1-p)^{K-a} p^{a+1} + o(p^{a+1})}) \\
	& \leq  (b (1-p)^{K-a} p^{a+1} + o(p^{a+1})) \\
	& \times \log(\frac{1}{b (1-p)^{K-a} p^{a+1} + o(p^{a+1})})  \\
	& \leq  (b (1-p)^{K-a} p^{a+1} + o(p^{a+1})) \\
	& \times \log(\frac{1}{ (1-p)^{K} p^{K} })  \\
	& =  K(b (1-p)^{K-a} p^{a+1} + o(p^{a+1})) \\
	& \times \log(\frac{1}{ p (1-p) }) \\
	& = o(p).
	\end{align*}

\subsection{Proof of Lemma \ref{lem:binary_scheme_performance}}

We will prove this by induction on $K \in \mathbb{N}.$  Let $p \in [0, 0.5].$  For $K = 1,$ we have that 
\begin{align*} 
& I(X \; ; \; X + Z_1) \\ 
& = H(p) - H(X \; | \; X + Z_1) \\
& = H(p) - 2p(1-p)H(0.5)
\end{align*}		
which matches the formula.
Assume the formula holds for the $(K-1)$th  case where $K > 1.$  Consider the $K$th case: 
\begin{align*} 
& I(X \; ; \; X + \sum_{i=1}^{K} 2^{i-1} Z_i) \\
& = H(p) - H(X \; | \; X + \sum_{i=1}^{K} 2^{i-1} Z_i) \\
& = H(p) - (1-p)H(X \; | \; X + \sum_{i=1}^{K-1} 2^{i-1} Z_i) \\
& - p H(X \; | \; X + \sum_{i=1}^{K-1} 2^{i-1} Z_i) \\
& - (p^{K}(1-p) + p(1-p)^{K}) H(\frac{p^{K} (1-p)}{p^{K} (1-p) + p (1-p)^{K}}) \\
& = H(p) - H(X \; | \; X + \sum_{i=1}^{K-1} 2^{i-1} Z_i) \\ 
& - (p^{K}(1-p) + p(1-p)^{K}) H(\frac{1}{1 + \frac{p(1-p)^K}{p^K(1-p)}}) \\
& = H(p) - \sum_{i=1}^K (p^i(1-p) + p(1-p)^i) H(\frac{1}{1 + \frac{p(1-p)^i}{p^i(1-p)}}).
\end{align*}

\subsection{Proof of Lemma \ref{lem:lower_bound}}

	Define $S_K = 1 + \sum_{i=1}^{K} \alpha_i.$   For any $K\in \mathbb{N},$ $p \in [0, 0.5],$ and $\alpha_i \in \mathbb{N}, \; i \in [K],$  we have that 
	\begin{align*} 
	& I(X \; ; \; X + \sum_{i=1}^{K} \alpha_i Z_i) \\
	& = H(p) - H(X \; | \; X + \sum_{i=1}^{K} \alpha_i Z_i) \\
	& \geq H(p) - (1 - p^{K+1} - (1-p)^{K+1}) \\
	& =  H(p) - 1 + p^{K+1} + (1-p)^{K+1}
	\end{align*}
	where the third line follows because
	\begin{align*} 
	& H(X \; | \; X + \sum_{i=1}^{K} \alpha_i Z_i) \\
	& = \sum_{t = -S_K}^{S_K} \Pr(X + \sum_{i=1}^{K} \alpha_i Z_i = t) H(X \; | \; X + \sum_{i=1}^{K} \alpha_i Z_i = t)  \\
	& = p^{K+1} \cdot 0 + (1-p)^{K+1} \cdot 0 \\ 
	& + \sum_{t = -S_K+2}^{S_K-2} \Pr(X + \sum_{i=1}^{K} \alpha_i Z_i = t) H(X \; | \; X + \sum_{i=1}^{K} \alpha_i Z_i = t) \\
	& \leq \sum_{t = -S_K+2}^{S_K-2} \Pr(X + \sum_{i=1}^{K} \alpha_i Z_i = t) \\
	& \leq 1 - p^{K+1} - (1-p)^{K+1}.
	\end{align*}

\subsection{Proof of Lemma \ref{lem:optimality_of_binary}}

For any $K \in \mathbb{N}$ and $p = 0.5,$ the performance of the binary scheme is given by
\begin{align*}
	& I(X \; ; \; X + \sum_{i=1}^{K} 2^{i-1} Z_i) \\ 
	& =  H(0.5) - 2\sum_{i=1}^K (0.5)^{i+1} H(0.5) \\
	& = 1 - \sum_{i=1}^K (0.5)^{i} \\
	& = 1 - (\frac{1 - (0.5)^{K+1}}{0.5} - 1) \\
	& = (0.5)^{K} \\
	& = 1 - 1 + (0.5)^{K+1} + (0.5)^{K+1}
\end{align*}
Thus, the performance of the binary scheme matches the lower bound for $p = 0.5.$

\subsection{Proof of Lemma \ref{lem:main_recursion}}
For $N \geq 2,$ we have that 
\begin{align*} 
& H(\tilde{X} \; | \; T_{M, N}) 
\\ & = 
\sum_{t=-S_{M,N}}^{S_{M,N}} \Pr(T_{M, N} = t) H(\tilde{X} \; | \; T_{M, N} = t) 
\\ & = 
\sum_{t=-S_{M,N}}^{-A_M-1} \Pr(T_{M, N} = t) H(\tilde{X} \; | \; T_{M, N} = t) 
\\ & + 
\sum_{t=A_M+1}^{S_{M,N}} \Pr(T_{M, N} = t) H(\tilde{X} \; | \; T_{M, N} = t) 
\\ & + 
\sum_{t=-A_M}^{A_M} \Pr(T_{M, N} = t) H(\tilde{X} \; | \; T_{M, N} = t) 
\\ & = \sum_{t=-S_{M,N-1}}^{S_{M,N-1}-2A_M-1} \! \! \!\!\!\!\!\!\!\! (1-p) \Pr(T_{M, N-1} = t) H(\tilde{X} \; | \; T_{M, N-1} = t)
\\ & + \sum_{t=-S_{M,N-1}+2A_M+1}^{S_{M,N-1}} \! \! \!\!\!\!\!\!\!\! p \Pr(T_{M, N-1} = t) H(\tilde{X} \; | \; T_{M, N-1} = t) 
\\ & + 
\sum_{t=-A_M}^{A_M} \Pr(T_{M, N} = t) H(\tilde{X} \; | \; T_{M, N} = t) 
\\ & = 
\sum_{t=-S_{M,N-1}+2A_M+1}^{S_{M,N-1}-2A_M-1} \Pr(T_{M, N-1} = t) H(\tilde{X} \; | \; T_{M, N-1} = t)
\\ & + 
\sum_{t=-S_{M,N-1}}^{-S_{M,N-1}+2A_M} \!\!\!\!  \!\!\!\!  (1-p) \Pr(T_{M, N-1} = t) H(\tilde{X} \; | \; T_{M, N-1} = t)
\\ & + 
\sum_{t=S_{M,N-1}-2A_M}^{S_{M,N-1}} p\Pr(T_{M, N-1} = t) H(\tilde{X} \; | \; T_{M, N-1} = t)
\\ & + 
\sum_{t=-A_M}^{A_M} \Pr(T_{M, N} = t) H(\tilde{X} \; | \; T_{M, N} = t) 
\\ & = 
H(\tilde{X} \; | \; T_{M, N-1}) 
\\ & - 
p \sum_{t=-S_{M,N-1}}^{-S_{M,N-1}+2A_M} \Pr(T_{M, N-1} = t) H(\tilde{X} \; | \; T_{M, N-1} = t)
\\ & - 
(1-p) \sum_{t=S_{M,N-1}-2A_M}^{S_{M,N-1}} \!\!\!\!\!\!\!\!  \Pr(T_{M, N-1} = t) H(\tilde{X} \; | \; T_{M, N-1} = t)
\\ & + 
\sum_{t=-A_M}^{A_M} \Pr(T_{M, N} = t) H(\tilde{X} \; | \; T_{M, N} = t)
\end{align*}
where the third equality follows from the fact that if $t < -A_M,$ 
\begin{align*} 
& H(\tilde{X} \; | \; T_{M, N} = t)
\\ & = 
H(\Pr(\tilde{X} = 1 \; | \; T_{M, N} = t))
\\ & = 
H\left(\frac{p\Pr(T_{M, N} = t \; | \; \tilde{X} = 1) }{\Pr(T_{M, N} = t)}\right)
\\ & = 
H\left(\frac{1}{1 + \frac{(1-p)\Pr(T_{M, N}-\tilde{X} = t +1)}{p\Pr(T_{M, N}-\tilde{X} = t -1)}}\right) 
\\ & = 
H\left(\frac{1}{1 + \frac{(1-p)^2\Pr(T_{M, N-1}-\tilde{X} = t +1+2^{N-1})}{(1-p)p\Pr(T_{M, N-1}-\tilde{X} = t -1+2^{N-1})}}\right) 
\\ & = 
H\left(\frac{1}{1 + \frac{(1-p)\Pr(T_{M, N-1}-\tilde{X} = t +1+2^{N-1})}{p\Pr(T_{M, N-1}-\tilde{X} = t -1+2^{N-1})}}\right) 
\\ & = 
H\left(\frac{p\Pr(T_{M, N-1} = t + 2^{N-1} \; | \; \tilde{X} = 1) }{\Pr(T_{M, N-1} = t + 2^{N-1})}\right)
\\ & =	 
H\left(\tilde{X} \; | \; T_{M, N-1} = t+ 2^{N-1}\right), 
\end{align*}
and if $t > A_M,$
\begin{align*}  
H(\tilde{X} \; | \; T_{M, N} = t) = 	H(\tilde{X} \; | \; T_{M, N-1} = t - 2^{N-1}).
\end{align*}

}

\end{document}